\documentclass[5p,twocolumn]{elsarticle}

\journal{Journal of Astroparticle Physics}
\pdfoutput=1 
\usepackage[]{siunitx} 
\usepackage{xspace}

\usepackage{xcolor}
\definecolor{darkred}{rgb}{0.5,0,0}
\definecolor{darkblue}{rgb}{0,0,0.5}
\definecolor{firebrick}{rgb}{0.75,0.125,0.125}
\definecolor{darkgreen}{rgb}{0,0.5,0}
\usepackage[colorlinks=true,linkcolor=firebrick,citecolor=darkgreen,urlcolor=darkblue]{hyperref} 
\usepackage{lineno}
\modulolinenumbers[5]

\usepackage{color}
\usepackage{tikz}
\usepackage{pgfplots}
\usepackage{pgf-umlsd}
\usepackage[]{fp}
\pgfplotsset{compat=1.17}
\usepackage{hhline} % For hhline (double line) 
\usepackage[percent]{overpic}
\usepackage{rotating}

\newcommand{\ARZhad}{\emph{ARZ2020 (had.)}\xspace}
\newcommand{\ARZem}{\emph{ARZ2020 (had. + EM)}\xspace}
\newcommand{\Alvarez}{\emph{Alvarez2009 (had.)}\xspace}

\begin{document}
\begin{frontmatter}

\title{Deep-learning-based reconstruction of the neutrino direction and energy for in-ice radio detectors}

\author[a]{C. Glaser}
\ead{christian.glaser@physics.uu.se}
\author[b]{S. McAleer}
\ead{smcaleer@uci.edu}
\author[a]{S. Stjärnholm}
\ead{Sigfrid.Stjarnholm.2141@student.uu.se}
\author[b]{P. Baldi}
\ead{pfbaldi@uci.edu}
\author[c]{S. W. Barwick}
\ead{sbarwick@uci.edu}

\address[a]{Uppsala University Department of Physics and Astronomy, Uppsala SE-75237, Sweden.}
\address[b]{Department of Information and Computer Science, University of California, Irvine, CA 92697, USA.}
\address[c]{Department of Physics and Astronomy, University of California, Irvine, CA 92697, USA.}

\begin{abstract}
Ultra-high-energy (UHE) neutrinos ($>10^{16}$~eV) can be measured cost-effectively using in-ice radio detection, which has been explored successfully in pilot arrays. A large radio detector is currently being constructed in Greenland with the potential to measure the first UHE neutrino, and an order-of-magnitude more sensitive detector is being planned with IceCube-Gen2.
For such shallow radio detector stations, we present an end-to-end reconstruction of the neutrino energy and direction using deep neural networks (DNNs) developed and tested on simulated data. 
The DNN determines the energy with a standard deviation of a factor of two around the true energy ($\sigma \approx$ 0.3 in $\log_{10}(E)$), which meets the science requirements of UHE neutrino detectors. For the first time, we are able to predict the neutrino direction precisely for all event topologies including the complicated electron neutrino charged-current ($\nu_e$-CC) interactions.
The obtained angular resolution shows a narrow peak at $\mathcal{O}$(\SI{1}{\degree}) with extended tails that push the 68\% quantile for non-$\nu_e$-CC (resp. $\nu_e$-CC interactions) to $\SI{4}{\degree} (\SI{5}{\degree})$.
This highlights the advantages of DNNs for modeling the complex correlations in radio detector data, thereby enabling measurement of neutrino energy and direction.

\end{abstract}
\end{frontmatter}
% \linenumbers
% \begin{document}
% \maketitle
% \flushbottom

\section{Introduction}
\label{sec:intro}
The detection of ultra-high-energy (UHE) neutrinos is a key to solving the 100-year-old mystery of the origin of cosmic rays and is one of the crucial milestones for astroparticle physics \cite{Astro2020NeutrinoAstronomy,Ackermann2022}. 
Their detection gives access to the most violent phenomena in the universe, those that happen in the vicinity of supermassive black holes (active galactic nuclei), in neutron star mergers, or in gamma-ray bursts. Furthermore, it allows for fundamental measurements of neutrino cross-sections and flavor ratios at energies beyond the reach of Earth-based accelerators like the LHC \cite{Astro2020Fundamental, Valera2022, EstebanCrossSection2022}.

A cost-efficient way to measure these UHE neutrinos above \SI{30}{PeV} of energy is via a sparse array of radio antenna stations installed, for instance, in the Arctic or Antarctic ice \cite{BarwickGlaserNu2022, ARALimit2020, COSPAR2019, CollaborationAartsenAbrahamEtAl2016, ARIANNALimit2020, Gen2WhitePaper, Gen2RadioICRC}: A neutrino interaction in the ice generates a few-nanoseconds-long radio flash that can be detected from kilometer-long distances due to the large attenuation length of radio signals in ice.  Because of the low expected flux, no UHE neutrino has been observed yet, but the technology has already been shown to work reliably with small test-bed arrays such as ARA and ARIANNA \cite{ARIANNALimit2020, ARALimit2020}. With the Radio Neutrino Observatory in Greenland (RNO-G) a much larger detector (35 stations) is being constructed at the moment \cite{RNOGWhitePaper2021} and an order-of-magnitude more sensitive radio detector is foreseen for IceCube-Gen2 \cite{Gen2WhitePaper, Gen2RadioICRC}.

With the first detection of a UHE neutrino on the horizon for the next years, the development of reconstruction methods becomes increasingly important. In addition, a good estimation of the energy and pointing resolution for different detector designs is crucial for planning IceCube-Gen2, which is happening at the moment. Two different station designs have been established: 
In a \emph{deep} design (as explored by ARA \cite{ARA}) antennas are placed into narrow boreholes down to a depth of up to \SI{200}{m}. This increases the sensitivity to neutrinos per detector station but also increases the costs per station and limits the choice of available antennas due to the narrow borehole. 
The second design is a \emph{shallow} detector station (as explored by ARIANNA \cite{ARIANNA}) with high-gain LPDA antennas installed a few meters below the surface. The Radio Neutrino Observatory in Greenland (RNO-G) combines both designs into hybrid detector stations. The radio detector of IceCube-Gen2 foresees a hybrid array of shallow-only stations interspersed with hybrid stations \cite{Gen2RadioICRC}.

\begin{figure}[t]
    \centering
    \includegraphics[width=.9\columnwidth]{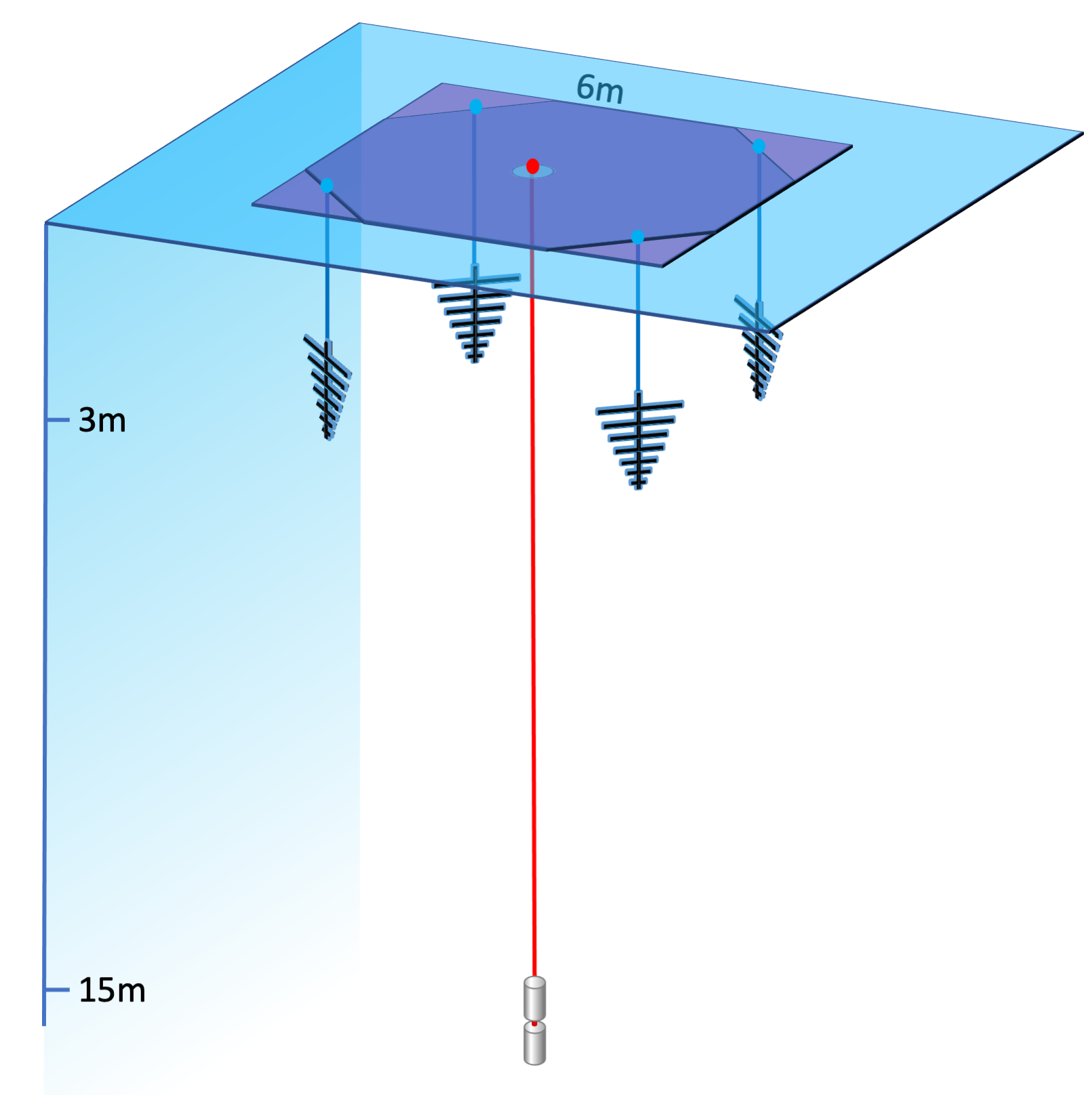}
    \caption{Schematic illustration of a shallow radio detector station. Fig. adapted from \cite{ARIANNA200}.}
    \label{fig:stationdesign}
\end{figure}

This work focuses on a shallow station design as shown in Fig.~\ref{fig:stationdesign}, which has been explored by the ARIANNA test-bed detector on the Ross Ice Shelf and at the South Pole \cite{ARIANNALimit2020}. Each station consists of 4 LPDA antennas installed at a depth of just a few meters below the snow surface, and 1 dipole antenna installed at a depth of \SI{10}{m} to \SI{15}{m} in a narrow borehole. These antennas observe the ice below for neutrino interactions. The dipole antenna was added to help with the reconstruction of the neutrino properties. At this depth, the antenna will observe two signals: one from a direct path to the antenna, and a second delayed signal from a reflection off the surface. The time difference between these two signals provides information about the distance to the neutrino interaction, which is important for estimating the neutrino energy \cite{DnR2019}. The exact depth is a compromise between the fraction of events that show a \emph{Direct and Reflected} (D'n'R) signature which decreases with depth (the shallower the better), the resolution of those events where a D'n'R signature is visible (the deeper the better), and deployment effort (the shallower the better) and is still being optimized. In the following, we set the depth to \SI{15}{m}; however, previous studies have indicated that \SI{10}{m} also will lead to energy resolutions that match the science requirements \cite{DnR2019}. In a future study, the work presented here can be used to study the impact of dipole depth on energy resolution in detail. 

Several aspects of the reconstruction of the neutrino direction and energy have been studied already and partly probed with in situ measurements. The reconstruction of the neutrino direction and energy requires measurement of the distance to the neutrino vertex, the signal arrival direction, the viewing angle, and the signal polarization, as well as a good understanding of the ice to correct for the bending of signal trajectories due to the changing index-of-refraction in the upper $\mathcal{O}$(\SI{100}{m}) of the ice sheet \cite{GlaserICRC2019}. For a shallow detector station, the measurement of the low-level parameters signal arrival direction (including the correction of ice propagation effects) and polarization has been probed experimentally at the South Pole using a transmitter lowered up to \SI{1.7}{km} deep into the ice \cite{GaswintPhD, ARIANNA:2020zrg, ARIANNA:2021pzm} as well as by measuring the radio emissions of cosmic rays \cite{ARIANNA2022CRPolarization}. In addition, a novel method for determining the vertex distance was developed and tested in an in situ measurement on the Ross Ice Shelf \cite{DnR2019}. This method uses the time difference between the direct and reflected-at-the-surface signals measured in the dipole. For a deep detector component, a similar method has been developed \cite{RNOGEnergy2021, ARADLICRC2021}.

To determine the neutrino direction at low signal-to-noise ratios, the \emph{forward folding} method was developed \cite{NuRadioReco}. Here an analytic model of the radio signal is fitted directly to the antennas' observed voltage measurements. The method has been applied successfully to shallow and deep detector components but requires a reconstruction of the neutrino interaction vertex as input \cite{GaswintPhD, ARIANNA:2021pzm, RNOGDirectionICRC2021}. In a Monte Carlo (MC) study using NuRadioMC \cite{NuRadioMC}, which includes a full trigger simulation and modeling of the signal chain and mixes the expected signals with noise, the angular resolution was determined to be approx. \SI{3}{\degree} for all triggered events of a shallow detector station \cite{GaswintPhD, ARIANNA:2021pzm}. For a deep detector, a similar resolution could only be obtained after a quality cut on the signal strength of the horizontally polarized antennas which reduced the fraction of usable events \cite{RNOGDirectionICRC2021}. A caveat of these analyses is that so far only hadronic particle cascades have been considered; these have a predictable shape and corresponding radio emission. For electron neutrino charged-current ($\nu_e$-CC) interactions, the generated electron generates an electromagnetic shower next to the hadronic shower. The electromagnetic shower is subject to the LPM effect so that for neutrino energies of approx. \SI{e18}{eV} and beyond, several spatially displaced subshowers are generated (see for instance the examples in \cite{NuRadioMC,BarwickGlaserNu2022}). At lower energies, often only the start of the shower is delayed, resulting in a displacement with respect to the  hadronic shower. Furthermore, the electromagnetic shower(s) can interfere with the hadronic shower. This can lead to more complex radio signal shapes but, more
importantly, the correlation between for instance the pulse width and the viewing angle changes,
which further complicates the reconstruction.
For $\nu_e$-CC interactions, no results on direction reconstruction have been reported so far because of the difficulty of modeling the LPM effect in traditional reconstruction methods \cite{GaswintPhD,RNOGDirectionICRC2021}.
Because $\nu_e$-CC interactions dominate the number of observable neutrinos\footnote{Assuming a 1:1:1 flavor ration, at \SI{e17}{eV} neutrino energy roughly 60\% of all triggered events stem from $\nu_e$-CC interactions. At neutrino energies of \SI{e18}{eV} the fraction reduces to 35\%.}, this paper addresses this significant caveat of current methods. We note that the signatures of $\nu_e$-CC interactions allow to distinguish them from non-$nu_e$-CC interactions. Preliminary results are promising \cite{ICRC2021DLdirection} and will be studied thoroughly in a forthcoming publication. In this work, we assume that the event type ($\nu_e$-CC or not) is known prior to the direction and energy reconstructions, but we also estimate the additional uncertainty resulting from a wrong guess of the event type.

For a deep detector station, an energy reconstruction was developed recently \cite{RNOGEnergy2021} with a shower energy resolution of 30\% for hadronic showers after moderate quality cuts. However, due to the differences in station design
between a deep and a shallow detector station,  these findings are not directly comparable to
our work. Here we focus on a shallow detector station, for which no end-to-end
reconstruction of the neutrino energy has been presented so far.

The reconstruction of the neutrino direction and energy from observable radio flashes is a complex problem: All information about the neutrino is compressed in a few-nanoseconds-long radio flash that is observed in just a few antennas. As a consequence, the development of traditional analysis approaches as described above is a very time-consuming process and the reconstruction algorithms often do not take into account all available information. In this article, we take a different approach: training  deep neural networks in a supervised manner \cite{baldi2021deep, erdmann2022} to extract the neutrino properties of interest directly from the simulated raw data. Deep-learning-based reconstructions have already produced promising results in closely related fields and have often outperformed existing methods, see e.g. \cite{Abbasi:2021ryj, PierreAuger:2021fkf}. We use the NuRadioMC code which simulates the expected signals from neutrinos \cite{NuRadioMC} as well as the corresponding detector response \cite{NuRadioReco} which allowed us to generate the needed training datasets. A preliminary version of the analysis was presented at the last International Cosmic Ray Conference (ICRC2021) \cite{ICRC2021DLenergy, ICRC2021DLdirection}.

The importance of the angular and energy resolution on the science output of in-ice radio detectors will depend strongly on the science case and also on the number of observed neutrinos which, in turn, depends on the largely unknown neutrino flux level. Different science cases and scenarios are currently under investigation. Recently, the potential to discover point sources \cite{Damiano2022} and to measure the neutrino-nucleon cross-section \cite{Valera2022} was presented for the future IceCube-Gen2 radio detector. Both analyses find a strong dependence on the angular resolution but only a weak dependence on the energy resolution. For an angular resolution of $\sigma =\,$ \SI{3}{\degree}\footnote{The authors quantify the angular resolution by quoting the uncertainty of the zenith angle only which is roughly $\sqrt{2}$ smaller than the space angle uncertainty quoted in this work if the zenith and azimuth uncertainties contribute equally to the total uncertainty.} promising results were obtained. Increasing the uncertainty to $\sigma =$ \SI{7}{\degree} (\SI{14}{\degree}) would increase the cross-section uncertainty by 50\% (178\%). For the point source study, increasing the angular resolution to $\sigma =$ \SI{7}{\degree} roughly doubles the size of the multiplets needed to claim discovery. The impact of energy resolution to distinguish different flux scenarios is currently under investigation.

\section{Dataset generation}

We use the NuRadioMC code \cite{NuRadioMC} to create a large dataset of around 40 million recorded neutrino events that correspond to a data volume of \SI{1.4}{TB}. We simulate a shallow radio detector station at the South Pole with the same configuration and trigger settings as foreseen for IceCube-Gen2 \cite{Gen2RadioICRC}: The station comprises two parallel pairs of downward-facing LPDA antennas \SI{2}{m} below the snow surface and an additional dipole antenna at a depth of \SI{15}{m} below the snow surface (see. Fig.~\ref{fig:stationdesign}). The detector is triggered by requiring a high and low-amplitude threshold crossing of the signal with an additional requirement of a time coincidence of two triggers out of the four LPDA antennas. The bandwidth of the trigger channels is reduced to 80-150~MHz to increase the sensitivity to neutrinos \cite{Glaser2020Bandwidth} but the full frequency information within 80-800~MHz is recorded for offline analysis. The trigger threshold is adjusted to yield a trigger rate on thermal noise fluctuations of \SI{100}{Hz}, as the data rate is the limiting factor in current and future experiments and thus sets the trigger threshold. We note that with a deep learning-based second-stage trigger that can run on the autonomous detector stations the data rate can be reduced significantly without impacting the neutrino efficiency \cite{ARIANNADLTrigger2021}.

\begin{figure}[t]
    \centering
    \includegraphics[width=1\columnwidth,  clip]{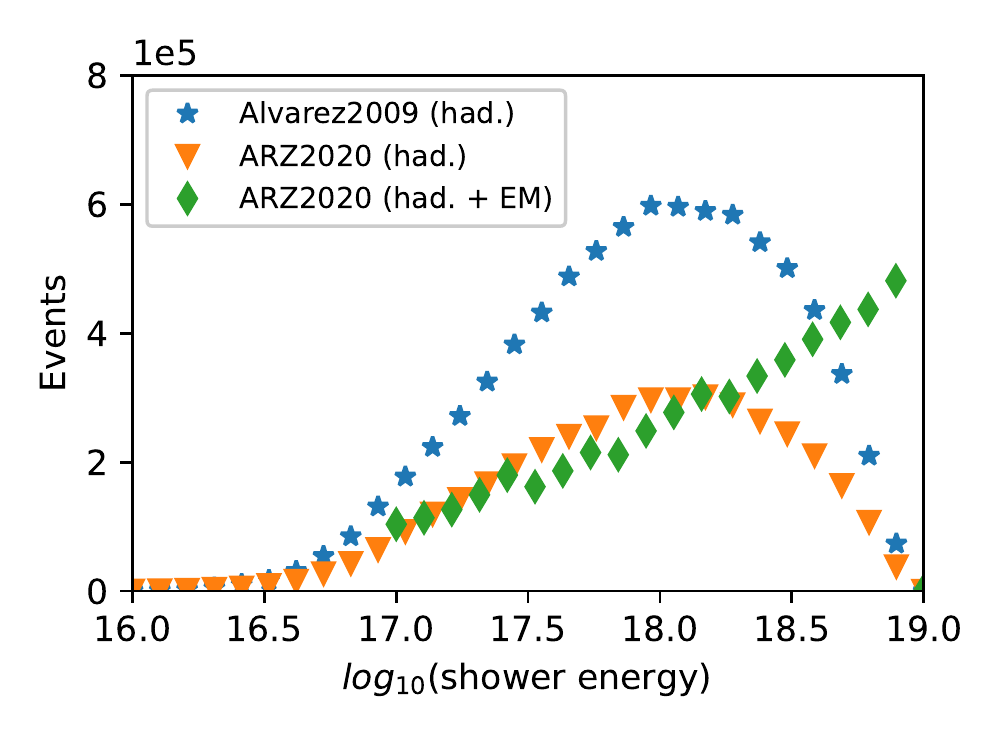}%
    \caption{Number of events in the 3 datasets as a function of the shower energy.}
    \label{fig:n_events}
\end{figure}

\begin{figure*}[t]
    \centering
    \begin{overpic}[width=.9\columnwidth%,grid,tics=10
        ]{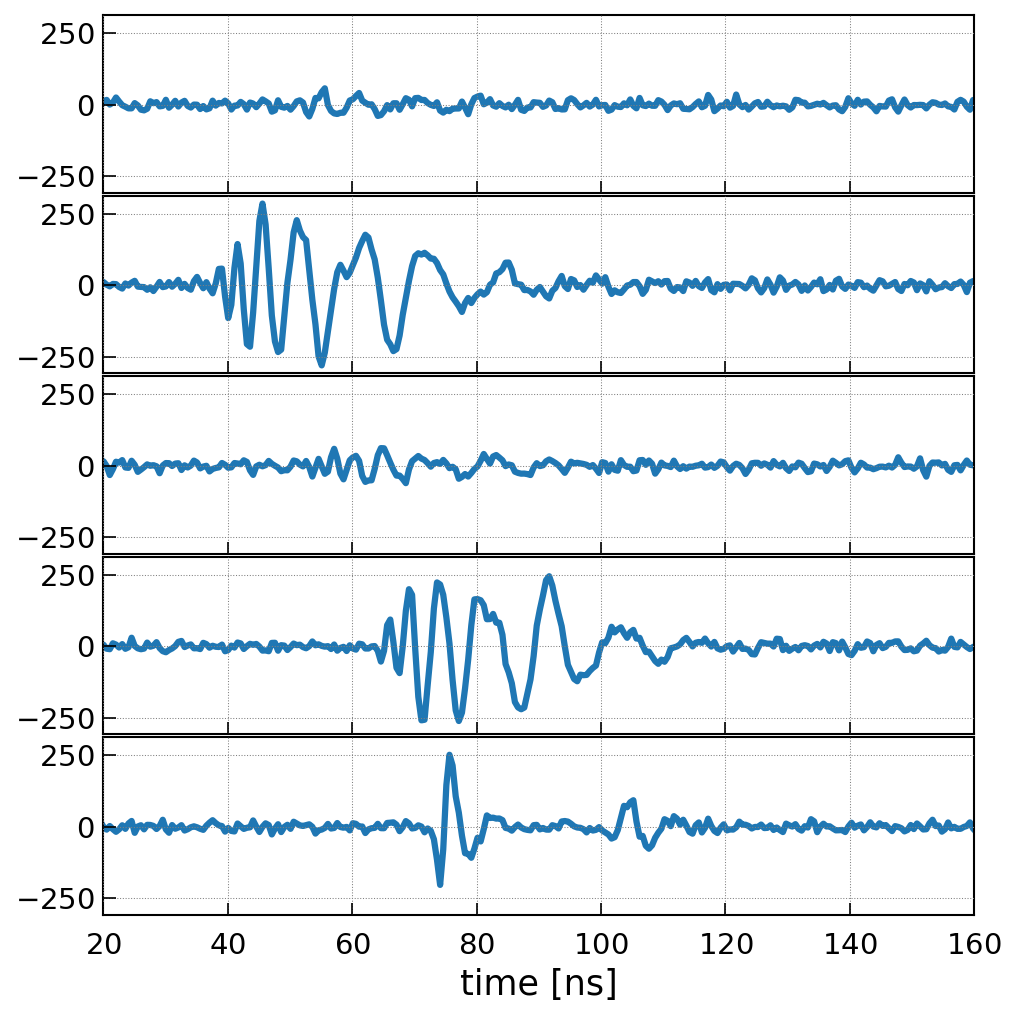}
        \put (0,40) {\begin{rotate}{90}voltage [$\mu$V]\end{rotate}}
    \end{overpic} 
    \includegraphics[width=.9\columnwidth]{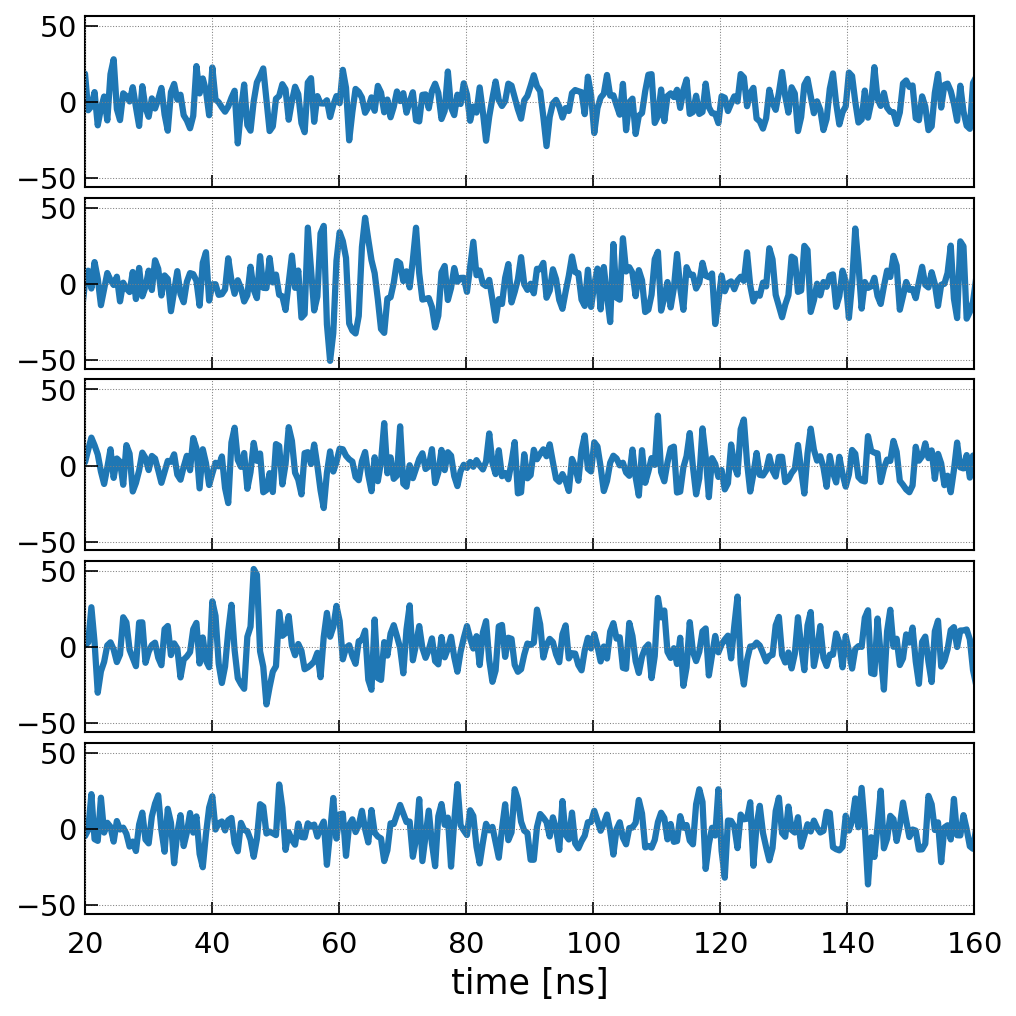}
    \caption{Two examples of the training dataset. The traces show the recorded voltages in the 5 antennas as a function of time. The upper 4 panels are the signal in the 4 LPDAs, and the bottom panel shows the signal in the dipole antenna. Left: example at a particularly high signal-to-noise ratio, where the direct and reflected pulse is clearly visible in the dipole antenna. Right: a more typical example at a low signal-to-noise ratio.}
    \label{fig:trace_example}
\end{figure*}

We generate neutrino interactions with neutrino energies ($E$) from \SI{e17}{eV} to \SI{e19}{eV}. The simulations are run separately for small energy bins (0.1 in $\log_{10}(E)$) in which the neutrino energies are distributed uniformly on a logarithmic scale. More simulations at low energies are executed to compensate for the lower trigger probability (see discussion below). The neutrino interactions are distributed uniformly in the ice around the detector with random incoming directions (uniformly distributed in the azimuth angle and the cosine of the zenith angle to assure a uniform sky coverage). We created three datasets with a comparable number of triggered events. The first dataset only contains hadronic showers -- i.e., particle cascades initiated by the breakup of the nucleus -- which are initiated by all neutrino interactions in the ice. The Askaryan signal is generated using the frequency domain parameterization \emph{Alvarez2009} \cite{Alvarez-Muniz2009, ZHAireS2012a} as implemented in NuRadioMC \cite{NuRadioMC}. This setup was used in previous analyses that used the forward folding technique \cite{GaswintPhD, RNOGDirectionICRC2021}. This dataset is referred to as \emph{Alvarez2009 (had.)} in the following. 

The second dataset also contains only hadronic showers but uses a more precise calculation of the Askaryan radiation: The \emph{ARZ2020} model calculates the Askaryan radiation from a library of charge-excess profiles of the particle cascades in the time domain \cite{NuRadioMC}, following the approach in \cite{ZHSTimeDomain, AlvarezMuniz2020}, which matches a microscopic MC simulation at a level of $\pm$3\%. This dataset is referred to as \emph{ARZ2020 (had.)} in the following. 

The third dataset contains electron neutrino charged-current ($\nu_e$-CC) interactions that produce both a hadronic shower as well as an additional electromagnetic shower through the outgoing electron. We again use the \emph{ARZ2020} model with the shower library available in NuRadioMC, which allows a proper simulation of the LPM effect that leads to a spatial delay of the electromagnetic shower with respect to the hadronic shower, and for increasing energy also to multiple spatially displaced sub-showers. We expect that this will complicate the reconstruction performance. Due to this additional complication, electron charged-current interactions were so far ignored in previous analyses. This dataset is referred to as \emph{ARZ2020 (had. + EM)} in the following.

The observed radio signals only depend on the shower properties. Therefore, we will train the deep neural network to determine the shower energy instead of the neutrino energy. The amount of energy transferred from the neutrino into the shower is a stochastic process and can be estimated from theory. For non-$\nu_e$-CC interactions, only a part of the neutrino energy is transferred to the particle shower. This stochastic process also introduces an irreducible uncertainty into the conversion from hadronic shower energy back to the neutrino energy which amounts to a standard uncertainty of 0.3 in $\log_{10}(E)$, i.e., a factor of two on a linear scale \cite{DnR2019}. This provides a target resolution for the shower energy. 
For $\nu_e$-CC interactions, the complete neutrino energy is deposited in the ice but the relative distribution into the hadronic and electromagnetic shower undergoes the same stochastic process. Hence, the shower energy equals the neutrino energy. We note that due to the LPM effect, the showers do not develop in phase (see discussion of the LPM effect above) which in practice often means that only a part of the shower energy is measured.

The likelihood of triggering the detector increases quickly with neutrino energy as the signal amplitude increases linearly with energy. We made an effort to simulate more low-energy neutrinos to obtain a similar number of triggered events across all energies, but our dataset is still biased toward high energies, as can be seen in Fig.~\ref{fig:n_events}. In this figure, we show the number of triggered events as a function of the shower energy which shows another interesting property. For the \emph{ARZ2020 (had. + EM)} dataset the shower energy equals the neutrino energy, and the number of triggered events steadily increases from \SI{e17}{eV} to \SI{e19}{eV}, as expected. In contrast, for the other two datasets, the distribution extends to lower energies because the shower energy is always lower than the neutrino energy. For the same reason, the number of events decreases at the high-energy end because it is unlikely that the complete neutrino energy is deposited into the hadronic shower. We expect that this uneven event distribution has a negative impact on the energy reconstruction at low energies due to an underrepresentation in the training dataset. 
In total the \emph{Alvarez2009 (had.)} dataset consists of \num{8.2e6} events, the \emph{ARZ2020 (had.)} dataset consists of \num{4.1e6} events, and the \emph{ARZ2020 (had. + EM)} consists of \num{5e6} events.

The training data consists of the recorded voltages as a function of time of the 5 antennas. We simulated a length of the time trace of \SI{256}{ns} with a time binning of \SI{0.5}{ns} which resulted in 512 samples per antenna. The shape of one event is $(5, 512)$. The data is normalized by using $\mu V$ as the base unit which brings the RMS of the data to $\mathcal{O}(10)$. The mean of the dataset is already zero, as the voltages fluctuate around the zero baseline. Two examples are shown in Fig.~\ref{fig:trace_example}.

\section{Direction Reconstruction}\label{s:direction_reconstruction}
\usetikzlibrary{matrix,patterns,spy,fit,calc}
\usepgfplotslibrary{groupplots}
\pgfplotsset{compat=newest}

\FPset{totalOffset}{0}

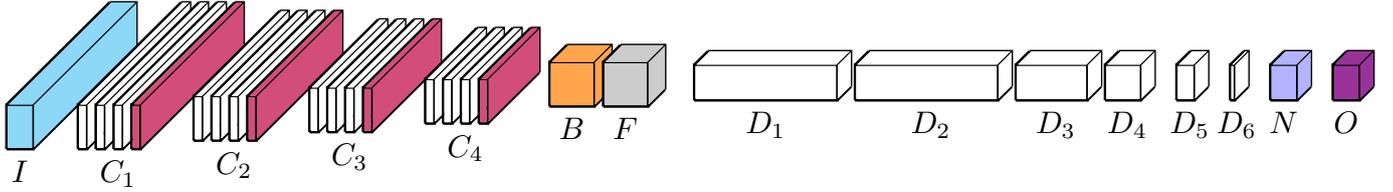
\begin{figure*}[t]
	\centering
	\noindent\resizebox{\textwidth}{!}{
    % \noindent\resizebox{400pt}{!}{
	\begin{tikzpicture}
		%\draw[use as bounding box, transparent] (-1.8,-1.8) rectangle (17.2, 3.2);

		\newcommand{\networkLayer}[9]{
			% Define the macro.
			% 1st argument: Height and width of the layer rectangle slice.
			% 2nd argument: Depth of the layer slice
			% 3rd argument: X Offset --> use it to offset layers from previously drawn layers.
			% 4th argument: Y Offset --> Use it when an output needs to be fed to multiple layers that are on the same X offset.
			% 5th argument: Z Offset --> Use to offset layers from previous 
			% 6th argument: Options for filldraw.
			% 7th argument: Text to be placed below this layer.
			% 8th argument: Name of coordinates. When name = "start" this resets the offset counter
			% REMOVED BY SIGFRID: THIS AGUMENT IS NO LONGER USED9th argument: list of nodes to connect to (previous layers)
			% 9th argument: hegith if not the same as width (set to 0 to have same width and height
			\xdef\totalOffset{\totalOffset}
 			\ifthenelse{\equal{#8} {start}}
 			{\FPset{totalOffset}{0}}
 			{}
 			\FPeval\currentOffset{0+(totalOffset)+(#3)}
            
            \ifthenelse{\equal{#9} {0}}
 			{
     			\def\h{#1} 
     			\def\w{#1}
 			}
 			{
 			    \def\h{#9} 
     			\def\w{#1}
 			}
			
			\def\b{0.02}
			\def\c{#2} % Width of the cube to distinguish number of input channels for current layer.
			\def\x{\currentOffset} % X offset for current layer.
			\def\y{#4} % Y offset for current layer.
			\def\z{#5} % Z offset for current layer.
			\def\inText{#7}

            % Define references to points on the cube surfaces
            \coordinate (#8_front) at  (\x+\c  , \z      , \y);
            \coordinate (#8_back) at   (\x     , \z      , \y);
            \coordinate (#8_top) at    (\x+\c/2, \z+\h/2, \y);
            \coordinate (#8_bottom) at (\x+\c/2, \z-\h/2, \y);
            
 			% Define cube coords
			\coordinate (blr) at (\c+\x,  -\h/2+\z,  -\w/2+\y); %back lower right
			\coordinate (bur) at (\c+\x,   \h/2+\z,  -\w/2+\y); %back upper right
			\coordinate (bul) at (0 +\x,   \h/2+\z,  -\w/2+\y); %back upper left
			\coordinate (fll) at (0 +\x,  -\h/2+\z,   \w/2+\y); %front lower left
			\coordinate (flr) at (\c+\x,  -\h/2+\z,   \w/2+\y); %front lower right
			\coordinate (fur) at (\c+\x,   \h/2+\z,   \w/2+\y); %front upper right
			\coordinate (ful) at (0 +\x,   \h/2+\z,   \w/2+\y); %front upper left

            % Draw connections from other points to the back of this node
            % REMOVED BY SIGFRID: THIS ARGUMENT IS NO LONGER USED.
            %         \ifthenelse{\equal{#9} {}}
 			% {}{
 			%     \foreach \val in #9
 			%         \draw[line width=0.3mm] (\val)--(#8_back);
 			% }
 			
			% Draw the layer body.
			% back plane
			\draw[line width=0.3mm](blr) -- (bur) -- (bul);
			% front plane
			\draw[line width=0.3mm](fll) -- (flr) node[midway,below] {\inText} -- (fur) -- (ful) -- (fll);
			\draw[line width=0.3mm](blr) -- (flr);
			\draw[line width=0.3mm](bur) -- (fur);
			\draw[line width=0.3mm](bul) -- (ful);

			% Recolor visible surfaces
			% front plane
			\filldraw[#6] ($(fll)+(\b,\b,0)$) -- ($(flr)+(-\b,\b,0)$) -- ($(fur)+(-\b,-\b,0)$) -- ($(ful)+(\b,-\b,0)$) -- ($(fll)+(\b,\b,0)$);
			\filldraw[#6] ($(ful)+(\b,0,-\b)$) -- ($(fur)+(-\b,0,-\b)$) -- ($(bur)+(-\b,0,\b)$) -- ($(bul)+(\b,0,\b)$);

			% Colored slice.
			\ifthenelse {\equal{#6} {}}{} % Do not draw colored slice if #4 is blank.
			% Else, draw a colored slice.
			{\filldraw[#6] ($(flr)+(0,\b,-\b)$) -- ($(blr)+(0,\b,\b)$) -- ($(bur)+(0,-\b,\b)$) -- ($(fur)+(0,-\b,-\b)$);}

			\FPeval\totalOffset{0+(currentOffset)+\c}
		}
		
	%\networkLayer{2.0}{0.5}{0.0}{0.0}{2.5}{color=red!50}{}{start}{}
	%\networkLayer{2.0}{0.25}{1.5}{0.0}{0.0}{color=green!50}{}{bot}{{start_front}}
	%\networkLayer{2.0}{0.25}{0.15}{0.0}{0.0}{color=green!50}{}{}{}
	%\networkLayer{2.0}{0.5}{0.15}{0.0}{0.0}{color=green!50}{}{end}{}
	%\networkLayer{2.0}{0.5}{-(2.8)/2}{0.0}{5.0}{color=green!50}{}{top}{{start_front}}
	%\networkLayer{2.0}{0.5}{2.0}{0.0}{2.5}{color=blue!50}{}{add}{{end_front,top_front}}
	%\networkLayer{1.0}{0.5}{0.15}{0.0}{2.5}{color=blue!50}{}{}{}
	%\networkLayer{0.75}{0.5}{0.15}{0.0}{2.5}{color=blue!50}{}{}{}
	%\networkLayer{0.5}{0.5}{0.15}{0.0}{2.5}{color=blue!50}{}{}{}
	
	    % Define the macro.
		% 1st argument: Height and width of the layer rectangle slice.
		% 2nd argument: Depth of the layer slice
		% 3rd argument: X Offset --> use it to offset layers from previously drawn layers.
		% 4th argument: Y Offset --> Use it when an output needs to be fed to multiple layers that are on the same X offset.
		% 5th argument: Z Offset --> Use to offset layers from previous 
		% 6th argument: Options for filldraw.
		% 7th argument: Text to be placed below this layer.
		% 8th argument: Name of coordinates. When name = "start" this resets the offset counter
		% 9th argument: list of nodes to connect to (previous layers)
	
	    \newcommand{\distanceintra}{0.1}
	    \newcommand{\distanceinter}{0.5}
	    \newcommand{\poolingcolor}{purple}
	    \newcommand{\poolingopacity}{70}
	
	    \newcommand{\smallthickness}{0.5}
	
	    \newcommand{\layersize}{3}
	
			% INPUT
		\networkLayer{\layersize}{0.3}{0.0}{0.0}{0.0}{color=cyan!40}{\(I\)}{start}{\smallthickness}

		% Conv block 1
		\networkLayer{\layersize}{0.1}{\distanceinter}{0.0}{0.0}{color=white}{}{}{\smallthickness}
		\networkLayer{\layersize}{0.1}{\distanceintra}{0.0}{0.0}{color=white}{}{}{\smallthickness}      
		\networkLayer{\layersize}{0.1}{\distanceintra}{0.0}{0.0}{color=white}{\(C_1\)}{}{\smallthickness}      
		% AveragePooling 
		\networkLayer{\layersize}{0.1}{\distanceintra}{0.0}{0.0}{color=\poolingcolor!\poolingopacity}{}{}{\smallthickness}        

        \renewcommand{\layersize}{2.5}
        
        % Conv block 2
		\networkLayer{\layersize}{0.1}{\distanceinter}{0.0}{0.0}{color=white}{}{}{\smallthickness}
		\networkLayer{\layersize}{0.1}{\distanceintra}{0.0}{0.0}{color=white}{}{}{\smallthickness}      
		\networkLayer{\layersize}{0.1}{\distanceintra}{0.0}{0.0}{color=white}{\(C_2\)}{}{\smallthickness}      
		% AveragePooling 
		\networkLayer{\layersize}{0.1}{\distanceintra}{0.0}{0.0}{color=\poolingcolor!\poolingopacity}{}{}{\smallthickness} 
		
		\renewcommand{\layersize}{2}
		
		% Conv block 3
		\networkLayer{\layersize}{0.1}{\distanceinter}{0.0}{0.0}{color=white}{}{}{\smallthickness}
		\networkLayer{\layersize}{0.1}{\distanceintra}{0.0}{0.0}{color=white}{}{}{\smallthickness}      
		\networkLayer{\layersize}{0.1}{\distanceintra}{0.0}{0.0}{color=white}{\(C_3\)}{}{\smallthickness}      
		% AveragePooling 
		\networkLayer{\layersize}{0.1}{\distanceintra}{0.0}{0.0}{color=\poolingcolor!\poolingopacity}{}{}{\smallthickness} 
		
		\renewcommand{\layersize}{1.5}
		
		% Conv block 4
		\networkLayer{\layersize}{0.1}{\distanceinter}{0.0}{0.0}{color=white}{}{}{\smallthickness}
		\networkLayer{\layersize}{0.1}{\distanceintra}{0.0}{0.0}{color=white}{}{}{\smallthickness}      
		\networkLayer{\layersize}{0.1}{\distanceintra}{0.0}{0.0}{color=white}{\(C_4\)}{}{\smallthickness}      
		% AveragePooling 
		\networkLayer{\layersize}{0.1}{\distanceintra}{0.0}{0.0}{color=\poolingcolor!\poolingopacity}{}{}{\smallthickness} 
        
        \renewcommand{\layersize}{0.5}
        
        % Batchnormalization & flatten
		\networkLayer{\layersize}{0.5}{\distanceinter}{0.0}{0.0}{color=orange!70}{\(B\)}{}{0}
		\networkLayer{\layersize}{0.5}{\distanceintra}{0.0}{0.0}{color=gray!40}{\(F\)}{}{0}
		
% 		\renewcommand{\layersize}{2.5}
%         \newcommand{\distanceintradense}{0.2}
        
%         % Dense
% 		\networkLayer{\layersize}{0.3}{\distanceinter}{0.0}{0.0}{color=white}{\(D_1\)}{}{}
% 		\networkLayer{\layersize}{0.3}{\distanceintradense}{0.0}{0.0}{color=white}{\(\;D_2\)}{}{}
		
% 		\renewcommand{\layersize}{2.0}
% 		\networkLayer{\layersize}{0.3}{\distanceintradense}{0.0}{0.0}{color=white}{\(\;D_3\)}{}{}
		
% 		\renewcommand{\layersize}{1.5}
% 		\networkLayer{\layersize}{0.3}{\distanceintradense}{0.0}{0.0}{color=white}{\(\;D_4\)}{}{}
        
%         \renewcommand{\layersize}{1.0}
% 		\networkLayer{\layersize}{0.3}{\distanceintradense}{0.0}{0.0}{color=white}{\(\;D_5\)}{}{}
		
% 		\renewcommand{\layersize}{0.5}
% 		\networkLayer{\layersize}{0.3}{\distanceintradense}{0.0}{0.0}{color=white}{\(\;D_6\)}{}{}

		\renewcommand{\layersize}{1.6}
        \newcommand{\distanceintradense}{0.2}
        \newcommand{\layerheightdense}{0.4}
        
        % Dense
		\networkLayer{\layerheightdense}{\layersize}{\distanceinter}{0.0}{0.0}{color=white}{\(D_1\)}{}{0}
		\networkLayer{\layerheightdense}{\layersize}{\distanceintradense}{0.0}{0.0}{color=white}{\(\;D_2\)}{}{0}
		
		\renewcommand{\layersize}{0.8}
		\networkLayer{\layerheightdense}{\layersize}{\distanceintradense}{0.0}{0.0}{color=white}{\(\;D_3\)}{}{0}
		
		\renewcommand{\layersize}{0.4}
		\networkLayer{\layerheightdense}{\layersize}{\distanceintradense}{0.0}{0.0}{color=white}{\(\;D_4\)}{}{0}
        
        \renewcommand{\layersize}{0.2}
		\networkLayer{\layerheightdense}{\layersize}{\distanceintradense*2}{0.0}{0.0}{color=white}{\(\;D_5\)}{}{0}
		
		\renewcommand{\layersize}{0.05}
		\networkLayer{\layerheightdense}{\layersize}{\distanceintradense*2}{0.0}{0.0}{color=white}{\(\;D_6\)}{}{0}
		
		\renewcommand{\distanceinter}{0.4}
		% normalization layer
		\networkLayer{\layerheightdense}{0.3}{\distanceinter}{0.0}{0.0}{color=blue!30}{\(N\)}{}{0}
        
		% OUTPUT
		\networkLayer{\layerheightdense}{0.3}{\distanceinter}{0.0}{0.0}{color=violet!80}{\(O\)}{}{0}          % Pixelwise segmentation with classes.

	\end{tikzpicture}
	}
	\caption{Sketch of the structure of the developed neural network. The labels are specified in Tab.~\ref{tab:method:cnn_model}. The relative sizes of the convolutional layers \(C_i\) hint at the layers' output dimensions, while the relative sizes of the dense layers \(D_i\) hint at the neuron amount in each layer.}
	\label{fig:methdod:cnn_model}
\end{figure*}

% Generated with     
% Nice code for plotting neural network things:
% https://davidstutz.de/illustrating-convolutional-neural-networks-in-latex-with-tikz/
% https://medium.com/momenton/typesetting-neural-network-diagrams-with-tex-4920b6b9fc19

In this work, we use a neural network architecture that follows the VGG model \cite{VGG2014}. The general idea of this architecture is the repeated application of convolutional layers followed by a single pooling, respectively. After a few of these blocks, the data is flattened and several fully connected layers are added to \emph{shrink} the information into the output nodes. For a detailed description of the different layers, the reader is referred to textbooks, e.g.,  \cite{baldi2021deep,erdmann2022}.  The VGG model was developed for image recognition. Similarly, one can interpret our data as one-dimensional images where the voltage as a function of time corresponds to the color amplitude as a function of spatial pixel position. 

The usage of convolutional layers matches the physical properties of the data. The pulse shape contains a lot of information about the neutrino properties. Traditionally, template matching techniques have been used to identify neutrino-induced radio pulses \cite{ARIANNALimit2020}. Convolutional layers can be thought of a more versatile and capable template where the subsequent application of small filters can match a multitude of different waveform shapes. This allows the neural network to identify and characterize the signal pulses that are invariant in time (cf. Fig.~\ref{fig:trace_example}). Because the signals' relative time and amplitude differences carry information, we apply the convolution filters independently to each antenna: i.e., we use an input shape of (5, 512, 1) corresponding to 5 antennas and 512 samples. The last component is 1 as the only available information from each antenna is the voltage amplitude (unlike three color channels in the case of images). We share the filter weights over the five antennas because the signal shapes are expected to be similar -- at least in the four LPDA antennas. A future improvement could be to have an individual set of filters for the dipole antenna, however, we expect the low-level features (e.g. sharp rising and falling amplitudes) to still be similar, independent of the antenna type.

An overview of the neural network architecture is presented in Fig.~\ref{fig:methdod:cnn_model}, with a more detailed description presented in Tab.~\ref{tab:method:cnn_model}. In the following, we first describe the best-performing model architecture to provide a good overview of the model before describing how we arrived at this particular model.

\begingroup
\setlength{\tabcolsep}{15pt}
\newcommand\Tstrut{\rule{0pt}{2.8ex}}         % = `top' strut
\newcommand\Bstrut{\rule[-1.5ex]{0pt}{0pt}}   % = `bottom' strut
\newcommand{\postblockspacing}{\Bstrut}
\newcommand{\preblockspacing}{\Tstrut}

\begin{table*}[tp]
\centering
\caption{Detailed structure of the model that was developed. The labels correspond to the labels used in Fig.~\ref{fig:methdod:cnn_model}. The dimensions specify the output dimensions of each layer, and the parameters specify the number of trainable parameters for each layer. Note that \textit{B} has 512 trainable parameters, but also 512 non-trainable parameters.}
\label{tab:method:cnn_model}
\begin{tabular}{llll}
\textbf{\textbf{Label}} & \textbf{\textbf{Layer}} & \textbf{\textbf{Dimensions}} & \textbf{\textbf{Parameters}} \\  \hhline{====}
\preblockspacing
\(I\)                   & Input layer             & \((5, 512, 1)\) & 0                   \postblockspacing \\ \hline
\preblockspacing
\(C_1\)                 & Conv2D (32 filters, kernel size (1,5))     & \((5, 512, 32)\) & 192                   \\
                        & Conv2D (32 filters, kernel size (1,5))     & \((5, 512, 32)\)   & 5 152                 \\
                        & Conv2D (32 filters, kernel size (1,5))     & \((5, 512, 32)\)    & 5 152                    \\
                        & AveragePooling2D        & \((5, 128, 32)\)   & 0                 \postblockspacing \\ \hline
\preblockspacing
\(C_2\)                 & Conv2D (64 filters, kernel size (1,5))     & \((5, 128, 64)\)  & 10 304                  \\
                        & Conv2D (64 filters, kernel size (1,5))     & \((5, 128, 64)\)   & 20 544                 \\
                        & Conv2D (64 filters, kernel size (1,5))     & \((5, 128, 64)\)     & 20 544               \\
                        & AveragePooling2D        & \((5, 32, 64)\)      & 0               \postblockspacing\\ \hline
\preblockspacing
\(C_3\)                 & Conv2D (128 filters, kernel size (1,5))    & \((5, 32, 128)\) & 41 088                   \\
                        & Conv2D (128 filters, kernel size (1,5))    & \((5, 32, 128)\) & 82 048                   \\
                        & Conv2D (128 filters, kernel size (1,5))    & \((5, 32, 128)\)  & 82 048                  \\
                        & AveragePooling2D        & \((5, 8, 128)\) & 0                    \postblockspacing\\ \hline
\preblockspacing
\(C_4\)                 & Conv2D (256 filters, kernel size (1,5))    & \((5, 8, 256)\)       & 164 096              \\
                        & Conv2D (256 filters, kernel size (1,5))    & \((5, 8, 256)\) & 327 936                    \\
                        & Conv2D (256 filters, kernel size (1,5) )    & \((5, 8, 256)\) & 327 936                    \\
                        & AveragePooling2D        & \((5, 2, 256)\)  & 0                   \postblockspacing\\ \hline
\preblockspacing
\(B\)                   & BatchNormalization      & \((5, 2, 256)\) & 512                    \postblockspacing\\ \hline \preblockspacing
\(F\)                   & Flatten                 & \((2560)\)    & 0                      \postblockspacing\\ \hline \preblockspacing
\(D_1\)                 & Dense (1024 units)      & \((1024)\)  & 2 622 464                        \\ 
\(D_2\)                 & Dense (1024 units)      & \((1024)\) & 1 049 600                         \\ 
\(D_3\)                 & Dense (512 units)       & \((512)\)   & 524 800                        \\ 
\(D_4\)                 & Dense (256 units)       & \((256)\)   & 131 328                        \\ 
\(D_5\)                 & Dense (128 units)       & \((128)\)    & 32 896                       \\ 
\(D_6\)                 & Dense (3 units)         & \((3)\)     & 387                        \postblockspacing\\ \hline \preblockspacing
\(N\)                   & Normalization           & \((3)\)    & 0                         \postblockspacing \\ \hline \preblockspacing
\(O\)             & Output layer          & \((3)\)  & 0                           \postblockspacing \\ \hline \preblockspacing
             & & \multicolumn{1}{r}{\textbf{Total:}} &  5 449 027
\end{tabular}
\end{table*}

\endgroup

The convolutional blocks consist of three convolutional layers, followed by pooling layers. The convolutional layers are configured to use padding, such that the dimensionality does not change following a convolutional layer. The filter size of the convolutional layers is (1, 5), which will result in the model scanning each antenna trace separately with the same filter. The average pooling layer has a pool size of 4, meaning it will decrease the trace's dimensions by a factor of 4 by taking the average of blocks of 4 samples. For example, in \(C_1\), the trace's length is reduced from 512 to 128 samples following the pooling layer. 
The number of filters is doubled in each block compared to the previous block. This is a very common technique in developing convolutional neural network models with pooling layers and can be seen as a way to mitigate underfitting due to the risk of the model becoming too simple if the dimensions decrease a lot, following the pooling layers. 

After four convolutional blocks, we apply a batch normalization and flatten the data tensor to a vector before applying a block of fully connected layers. 
The \emph{dense} block consists of 6 fully connected layers. The size of the fully connected layers decreases for layers further along the network and reduces the network to the 3 output neurons. The fully connected layers compile the information gathered by the convolutional layers, especially in regard to the time difference between the pulses in the 5 antenna traces.

We encode the neutrino direction as a three-dimensional cartesian unit vector that points in the direction of the neutrino origin. This is often the better choice compared to the zenith and azimuth angle in spherical coordinates due to the correlation between the two angles and the singularities at the poles. Therefore, we add an L2-normalization layer to make sure that the network prediction is a unit vector.

 \subsection{Training and Optimization of Network Architecture}
\label{sec:training}

We use tensorflow/Keras to train the neural network \cite{tensorflow2015-whitepaper, chollet2015keras}.
We reserved 300,000 events for an independent test dataset. We used 87\% of the remaining data for the training dataset and 13\% for the validation dataset. The datasets \emph{Alvarez2009 (had.)}, \emph{ARZ2020 (had.)}, and \emph{ARZ2020 (had. + EM)} contained  \num{8.2e6} events,  \num{4.1e6} events, and  \num{5e6} events respectively. Each event has a size of \SI{20}{kB} resulting in data volumes between \SI{82}{GB} and \SI{160}{GB} per dataset. The neural network is trained independently for each data set. The training dataset is so large that it does not fit into memory. To avoid slowing down the training process due to I/O and data preprocessing times, we developed a data pipeline based on the \emph{tensorflow.data} class, which preloads and preprocesses the data on several cores. We used a batch size of 64 and ran the training on an NVIDIA Quadro RTX 6000 GPU resulting in in a training time of roughly \SI{10}{min} per epoch. We always made sure that the training converged and that no overfitting took place which took a few hours of training time.

\begin{figure*}[t]
\centering
\includegraphics[width=0.3\textwidth, trim={0.3cm 0 0.3cm 0}, clip]{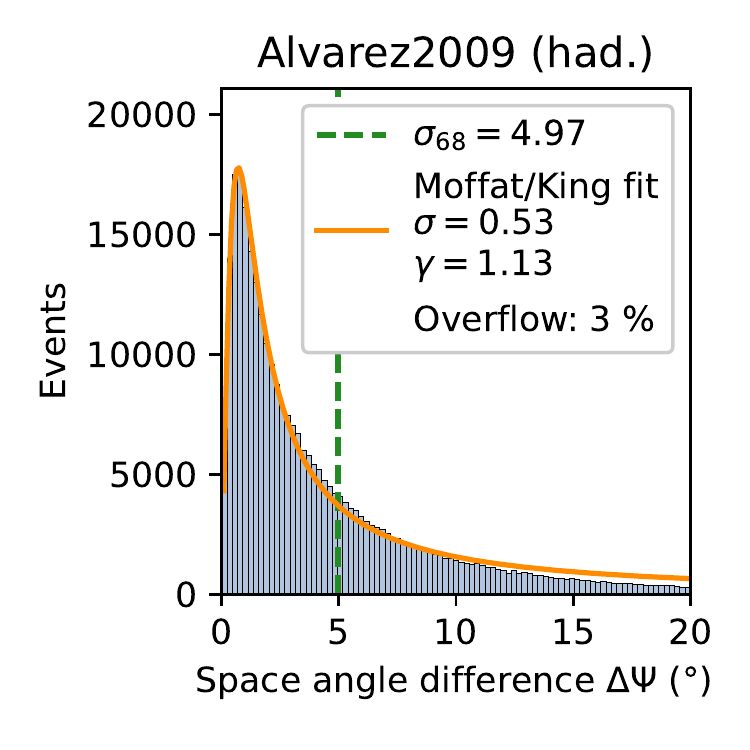}
\includegraphics[width=0.3\textwidth, trim={0.3cm 0 0.3cm 0}, clip]{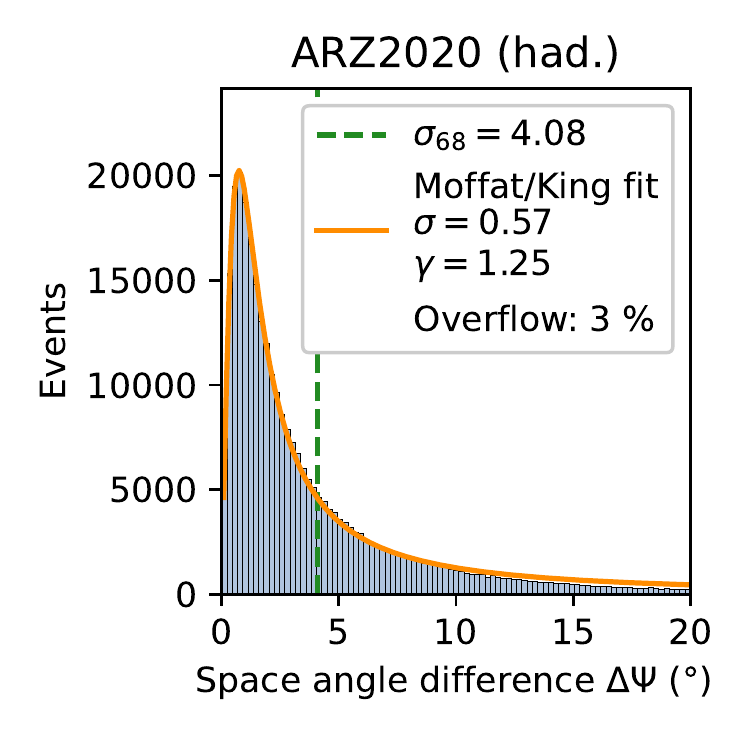}
\includegraphics[width=0.3\textwidth, trim={0.3cm 0 0.3cm 0}, clip]{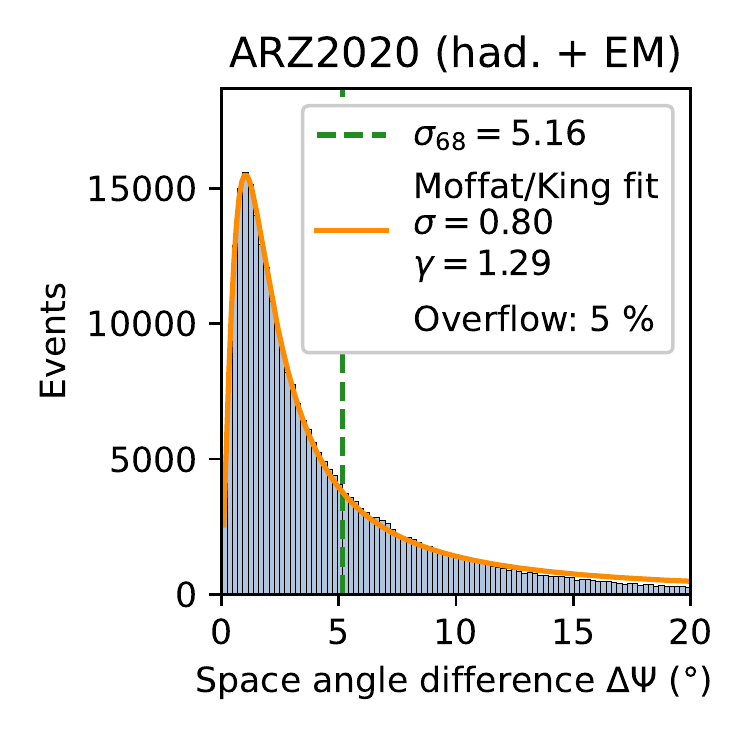} 
\caption{Space angle difference for the three datasets. A Moffat/King function gives a good description of the point spread function. The vertical dashed line shows the position of the 68\% quantile. The overflow in the legend specifies the fraction of events that have a space angle difference larger than \SI{20}{\degree}.}
\label{fig:res_resolution_histogram}
\end{figure*}

To train the model, we use the optimizer \textit{Adam} \cite{kingma2017adam} with a learning rate of \(5 \cdot 10^{-5}\) and the mean absolute error (MAE) error as loss function. The latter choice is motivated by the expectation that a certain percentage of the events will have too low signal quality to enable a good reconstruction. The  mean-squared error (MSE) loss will put more weight on events that might intrinsically be harder to reconstruct. We do not want these events to dominate the training. In a future analysis, quality cuts will be developed to remove those events from the dataset. We found that for this particular task of direction reconstruction, the MAE loss performed better than the MSE loss.

Through many experiments, we tuned the network inference performance via an iterative procedure by testing different hyperparameters but staying within the general VGG-like architecture of convolutional + pooling blocks followed by fully connected layers. 
The best-performing network was set as the one to be compared against for further experiments. During the optimizations, model parameters such as learning rate, number of convolutional and fully connected layers, activation functions, batch normalization use, loss functions, choice of the optimizer as well as convolutional layer stride length, filter size, and filter amount were tested in order to find the best-performing network structure and hyperparameters. We tested the parameters multiple times during the optimization to make sure that the values were not derived from a previous minimum that had moved when other parameters were tuned. When additional experimentation was performed with little or no improvements to the model performance, the optimization of the network was done. We considered other factors as well such as the number of trainable model parameters, where a model with a smaller number of trainable parameters was chosen when deciding between models that had equal performance. This made the model less complex, while also decreasing the required time for training and inference.

\subsection{Reconstruction resolution and dependence on event properties}
In the following, we show the results of the direction reconstruction for the three different datasets \emph{Alvarez2009 (had.)}, \emph{ARZ2020 (had.)}, and \emph{ARZ2020 (had. + EM)}. All results are evaluated for the independent test data set of 300,000 events (cf. Sec.~\ref{sec:training}).
Fig.~\ref{fig:res_resolution_histogram} shows histograms of the space angle difference \(\Delta\Psi\) between the true and reconstructed neutrino direction. We quantify the angular resolution by quoting the position of the 68\% quantile \(\sigma_{68}\) which is also specified in the graphs. 
 The point spread function can be described with a Moffat/King function that can be thought of as a Gaussian with extended tails \cite{https://doi.org/10.48550/arxiv.2202.11120, Moffat,King} and is given by:
\begin{equation} \label{eq:res_moffatking_fit_fcn}
    f(x; A, \sigma, \gamma) = A \frac{x}{2\pi\sigma^2}\left ( 1 - \frac{1}{\gamma} \right ) \left [ 1 + \frac{1}{2\gamma} \frac{x^2}{\sigma^2}\right ]^{-\gamma}
\end{equation}

We find that the narrow part of the distribution is described with a sigma of \SI{0.5}{\degree} to \SI{0.6}{\degree} for the hadronic datasets which increases to \SI{0.8}{\degree} for the more challenging $\nu_e$-CC dataset. Due to the tails of the distribution, the 68\% quantiles are at \SI{4}{\degree} to \SI{5}{\degree}. This finding highlights the advantage of a DNN for event reconstruction. The complex correlations between neutrino direction and observed radio flashes of $\nu_e$-CC interactions are learned well and are only slightly worse than for the simpler non-$\nu_e$-CC interactions. This is a significant improvement over previous reconstruction methods that fail to reconstruct $\nu_e$-CC interactions \cite{RNOGDirectionICRC2021,GaswintPhD}.

\begin{figure}[t]
    \centering
    \includegraphics[width=0.9\columnwidth, clip]{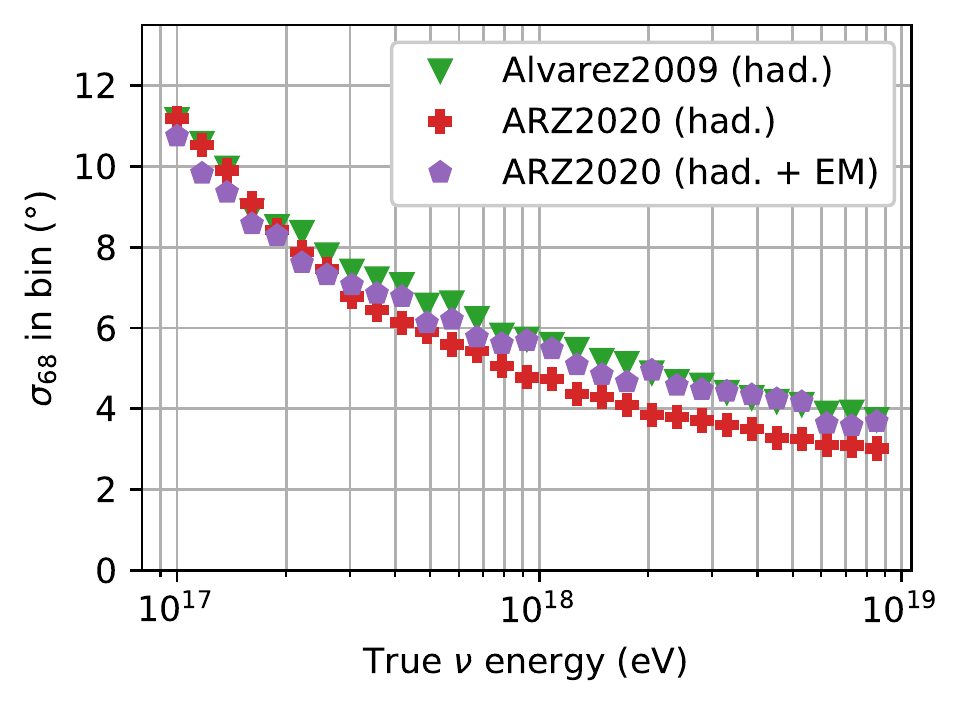}
    \caption{Angular resolution \(\sigma_{68}\) as a function of neutrino energy for the three datasets.}
    \label{fig:res_resolution_energy}
\end{figure}

\begin{figure}[t]
    \centering
    \includegraphics[width=0.9\columnwidth]{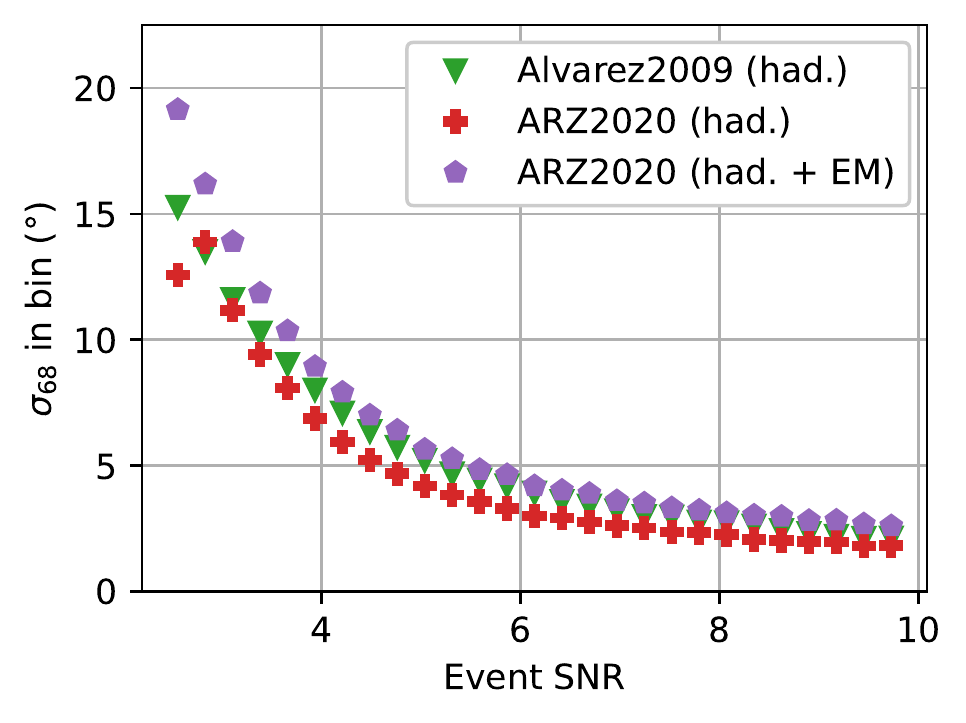}
    \caption{Angular resolution \(\sigma_{68}\) as a function of signal-to-noise ratio for the three datasets.}
    \label{fig:res_resolution_SNR}
\end{figure}

\begin{figure}[t]
    \centering
    \includegraphics[width=0.99\columnwidth]{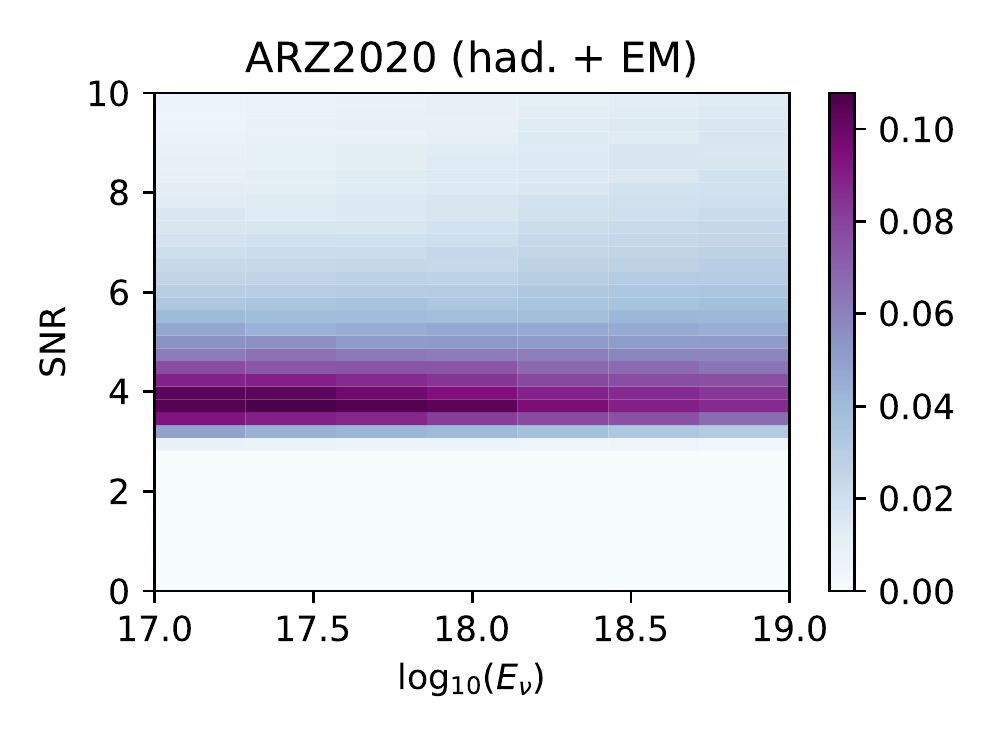}
    \caption{Correlation between neutrino energy and signal-to-noise ratio. Each column is normalized to 1. Here, the test data set of the ARZ2020(had. + EM) model is shown that is also representative for the other datasets. }
    \label{fig:correlation_SNR_energy}
\end{figure}

The average resolution strongly depends on the event distribution, in particular the number of events as a function of energy (cf. Fig.~\ref{fig:n_events}). Therefore, the average resolution obtained on the test data set needs to be interpreted with caution. A better way of showing the resolution is by quoting the resolution in bins of neutrino energy and/or signal-to-noise ratio (SNR) which is presented in Figs.~\ref{fig:res_resolution_energy} and \ref{fig:res_resolution_SNR}. We define the SNR of an event as the maximum of the absolute value of the LPDA traces, divided by the noise floor, which is approximately \SI{10}{\mu V}. The maximum signal amplitude is determined from noisy traces which results in typical SNR values of 3.5 or higher. Smaller values are rare. We observe an increase in performance with increasing energy and SNR as expected. The SNR and energy is weakly correlated with higher-energy events having more often high SNRs than low-energy events. We show the correlation in Fig.~\ref{fig:correlation_SNR_energy}.  However, in each energy bin, the SNR distribution still peaks at low SNRs, i.e., most events are still measured with signals close to the trigger threshold independent of the neutrino energy. Therefore, the improvement with energy can't be solely explained by an increasing SNR.

The three datasets' energy dependence gives interesting insights. Overall, the \emph{ARZ2020 (had.)} dataset performs best, especially at higher energies. The \emph{Alvarez2009 (had.)} dataset performs worse at almost all energies which was a surprise at first. Our initial expectation was a better performance because the simpler \emph{Alvarez2009} emission model should be easier to model which is indeed the case for traditional reconstruction techniques \cite{GaswintPhD, RNOGDirectionICRC2021}. We speculate that the reason for the better performance of the \emph{ARZ2020} emission model is that it generates more information than the neural network is able to learn. The \emph{Alvarez2009} model corresponds to a frequency domain parameterization where the phase was assumed to be a constant \SI{90}{\degree} at all frequencies leading to perfectly symmetric pulse forms. However, a more precise calculation of the radio emission in the time domain (as done in \emph{ARZ2020}) shows additional structures in the pulse shape that depend on the even geometry (see. e.g. \cite{AlvarezMuniz2020,NuRadioMC,ZHSTimeDomain}). 
This highlights the advantages of deep neural networks for event reconstruction.

The \emph{ARZ2020 (had. + EM)} dataset performs slightly worse at larger neutrino energies where the LPM effect becomes relevant and the stochasticity in the shower development complicates the reconstruction. Nevertheless, the resolution is just one to two degrees worse than for the simpler hadronic showers which is encouraging as traditional reconstruction techniques failed to reconstruct this event class with good resolution \cite{RNOGDirectionICRC2021,GaswintPhD}. At low energies where the LPM effect is negligible, the performance is similar to the \emph{ARZ2020 (had.)} dataset. The lowest energy bins are difficult to interpret because of low statistics in these bins.   
Another factor that influences that dependence on the energy is the different statistics in the training dataset. Part of the reason for the worse performance at lower energies might just be an underrepresentation in the training dataset.

The resolution as a function of the signal-to-noise ratio of Fig.~\ref{fig:res_resolution_SNR} shows a similar trend. Ignoring the first bin which has poor statistics, the resolution improves with increasing SNR. The performance of the \emph{ARZ2020 (had.)} dataset is the best over the entire range of SNR values, followed by the \emph{Alvarez2009 (had.)} dataset and the \emph{ARZ2020 (had. + EM)} dataset.


\begin{figure}[t]
    \centering
    \includegraphics[width=0.99\columnwidth]{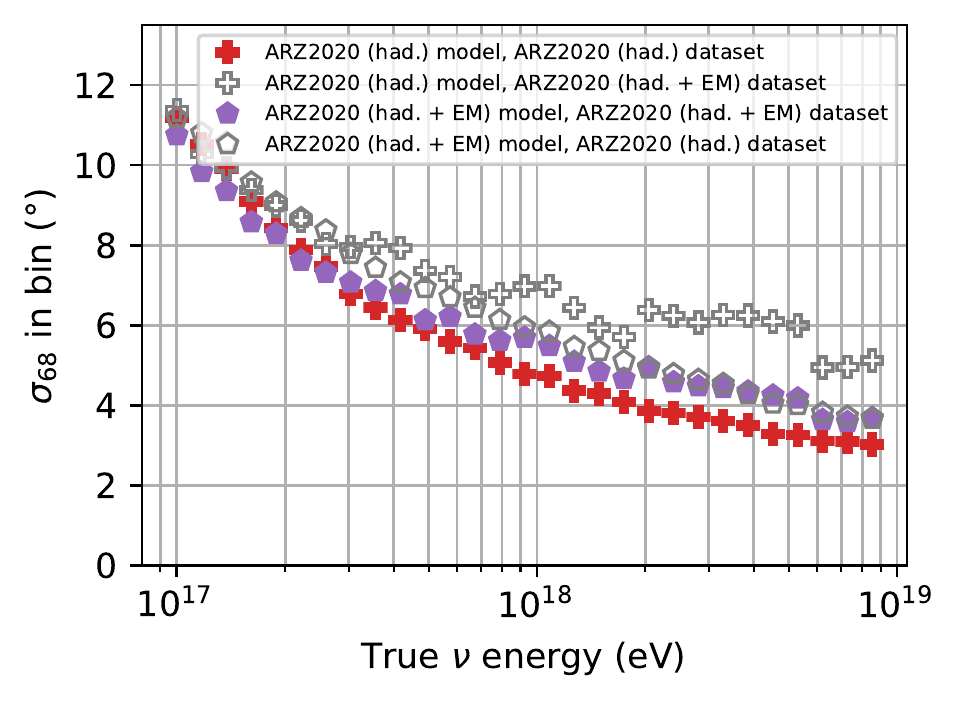} 
    \caption{Angular resolution \(\sigma_{68}\) as a function of neutrino energy for an incorrect guess of the event type, i.e., the neural network trained on $\nu_e$-CC events is evaluated on the non-$\nu_e$-CC dataset and vice versa. Filled markers show the correct combinations, empty markers the incorrect combinations. }
    \label{fig:ang_res_wrong_topology}
\end{figure}

So far we assumed that we know the event type, i.e., we reconstructed $\nu_e$-CC events with the neural network trained on $\nu_e$-CC events, and the same for non $\nu_e$-CC events. The signature of $\nu_e$-CC interactions allows to distinguish them from non-$\nu_e$-CC interactions, \cite{ICRC2021DLdirection}, however, with uncertainties. Therefore, as we do not always know the event type, we also studied the deterioration in  reconstruction performance when the \textit{wrong} network is used for reconstruction. The results are presented in Fig.~\ref{fig:ang_res_wrong_topology}. We observe that the model trained on $\nu_e$-CC events performs better on non-$\nu_e$-CC events than the other way around. This is expected because the $\nu_e$-CC dataset also contains events that are dominated by the hadronic shower and, thus, are similar to the  non-$\nu_e$-CC dataset. The  \emph{ARZ2020 (had. + EM)} network reconstructs the  \emph{ARZ2020 (had.)} dataset almost as good as the  \emph{ARZ2020 (had. + EM)} dataset it was trained on, with a similar performance above $E_\nu > $\SI{e18}{eV} and a deterioration of approx.~\SI{1}{\degree} for smaller energies. However, the \emph{ARZ2020 (had.)} network still performs approx.~\SI{1}{\degree} better on the \emph{ARZ2020 (had.)} dataset over the full energy range. 
Thus, if the event type is now known, it is best to use the \ARZem network but it will lead to a deterioration in angular resolution of approx.~\SI{1}{\degree} for non-$\nu_e$ interactions.

\begin{figure}[t]
    \centering
    \includegraphics[width=0.99\columnwidth]{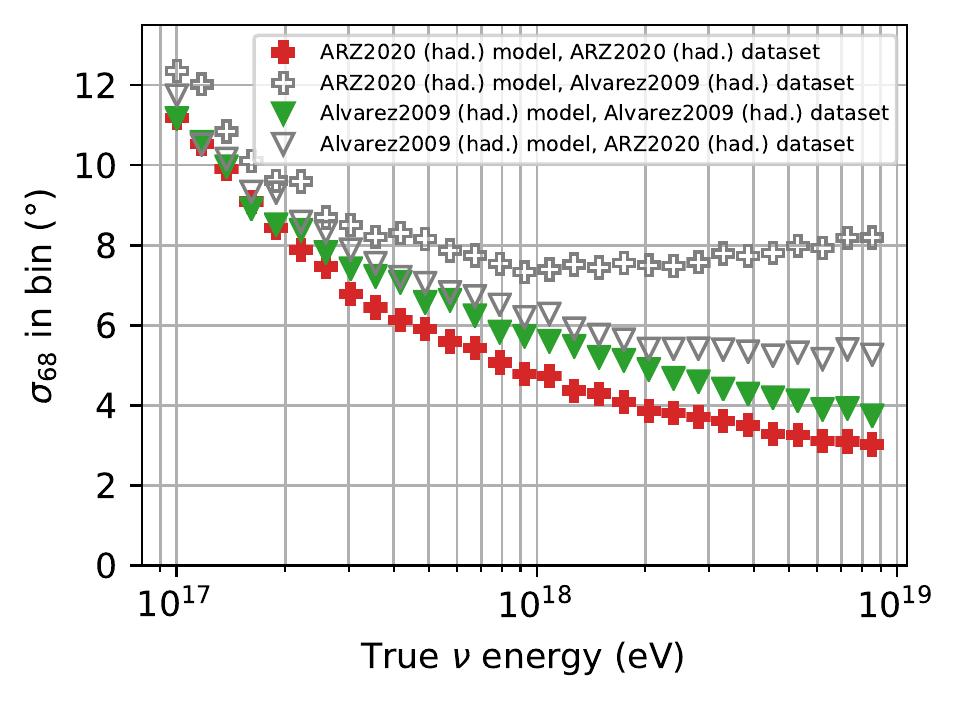} 
    \caption{Angular resolution \(\sigma_{68}\) as a function of neutrino energy for an incorrect combination of the Askaryan emission model, i.e., the neural network trained on the \emph{Alvarez2009 (had.)} dataset is evaluated with the \emph{ARZ2020 (had.)} dataset and vice versa. Filled markers show the correct combinations, empty markers the incorrect combinations. }
    \label{fig:ang_res_wrong_askaryan}
\end{figure}

We can also obtain an estimate of the systematic uncertainty of the Askaryan emission model by analyzing the \emph{Alvarez2009 (had.)} dataset with the network trained on the \emph{ARZ2020 (had.)} dataset and vice versa. Because we know that the \emph{ARZ2020} model is more accurate, the deterioration of the angular resolution does not represent the systematic uncertainty of the emission model but rather sets an upper bound.  The result is presented in Fig.~\ref{fig:ang_res_wrong_askaryan}. We find that the \emph{Alvarez2009} network works better on the \emph{ARZ2020} dataset than the other way around. A possible explanation for this behavior is that the \emph{Alvarez2009} network only learned simple features that are also present in the \emph{ARZ2020} dataset. In contrast, the \emph{ARZ2020} network learned to use second order features that give it a better performance than the \emph{Alvarez2009} network (see discussion above and Fig.~\ref{fig:res_resolution_energy} and \ref{fig:res_resolution_SNR}) but as these features are not present in the \emph{Alvarez2009} dataset it performs worse when applied to this dataset.

\subsection{Comparison to previous results}
A comparison with the work of \cite{GaswintPhD, ARIANNA:2021pzm} indicates that our DNN reconstruction still has room for improvement. The work of \cite{GaswintPhD, ARIANNA:2021pzm} used the forward folding technique \cite{NuRadioReco} and found a \SI{3}{\degree} resolution for hadronic showers (\emph{Alvarez2009 (had.)}) with little energy dependence. The analysis however assumed that the neutrino vertex position was known and only simulated uncertainties of the core position along the signal trajectory, whereas our DNN uses only the raw data without any additional information. But as this is a reasonable assumption, we expect that the limit of the DNN reconstruction is not yet reached and can be further improved in future work. 
For $\nu_e$-CC interactions, no results on direction reconstruction have been reported so far because of the difficulty of modeling the LPM effect in traditional reconstruction methods \cite{GaswintPhD,RNOGDirectionICRC2021}.

\section{Energy Reconstruction}

The structure of the neural network used for energy reconstruction is very similar to the one used for direction reconstruction, with the most notable difference of only one output node representing the reconstructed shower energy. In all other regards, the models used for direction and energy reconstruction are identical, which illustrates the benefits of using machine learning for these kinds of tasks. We perform the training in the logarithm of the shower energy, we use the mean squared error (MSE) as the objective function and otherwise use the same settings as before. In addition, we utilized the \textit{ReduceLROnPlateau} callback in Keras \cite{ReduceLROnPlateau} to reduce the learning rate when the loss stagnated.

\begin{figure*}[t]
    \centering
    \includegraphics[width=0.32\textwidth, trim={0.3cm 0 0.3cm 0},  clip]{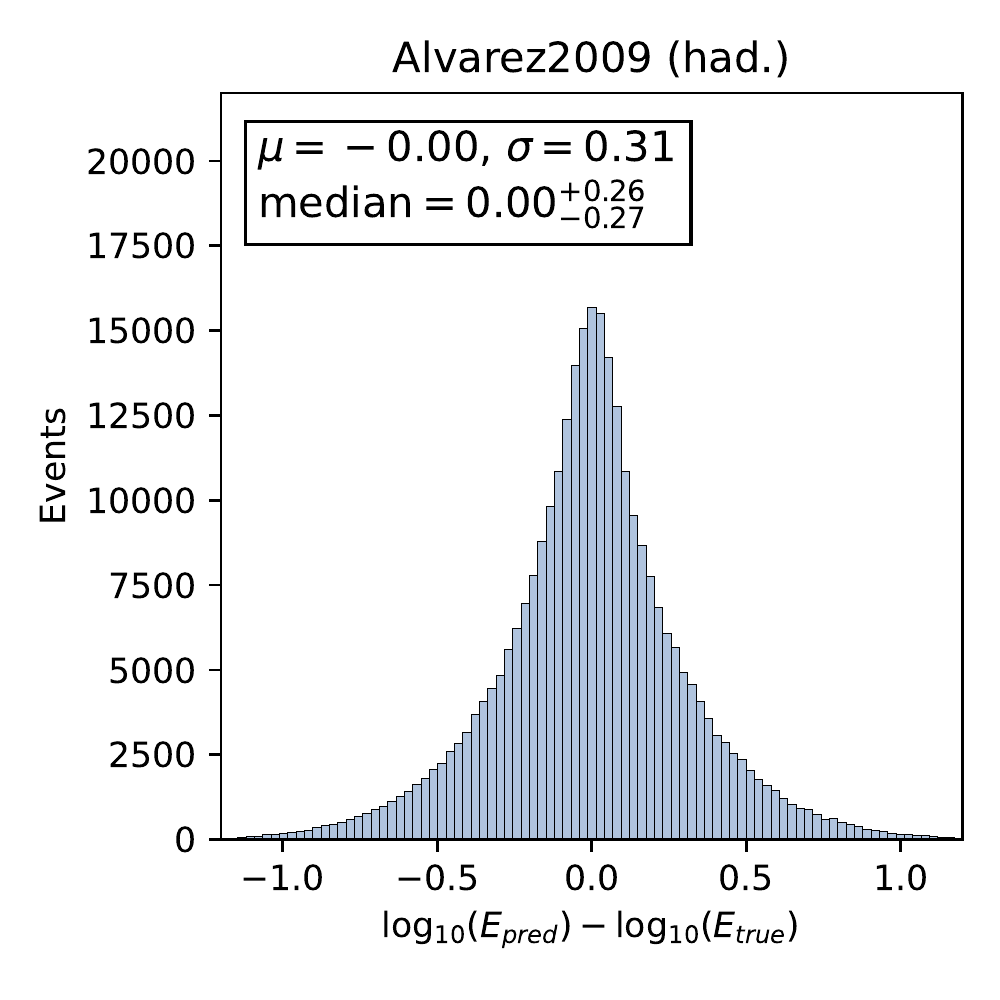}
    \includegraphics[width=0.32\textwidth, trim={0.3cm 0 0.3cm 0}, clip]{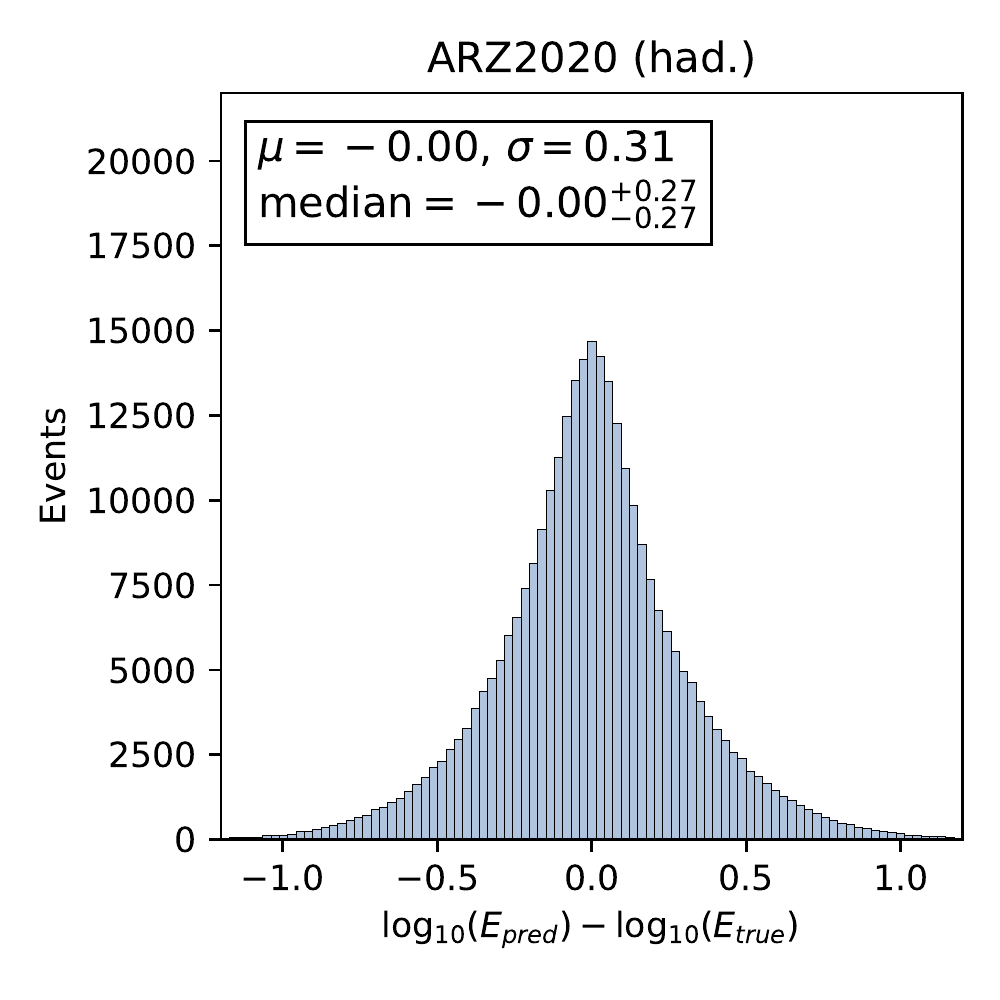}
    \includegraphics[width=0.32\textwidth, trim={0.3cm 0 0.3cm 0}, clip]{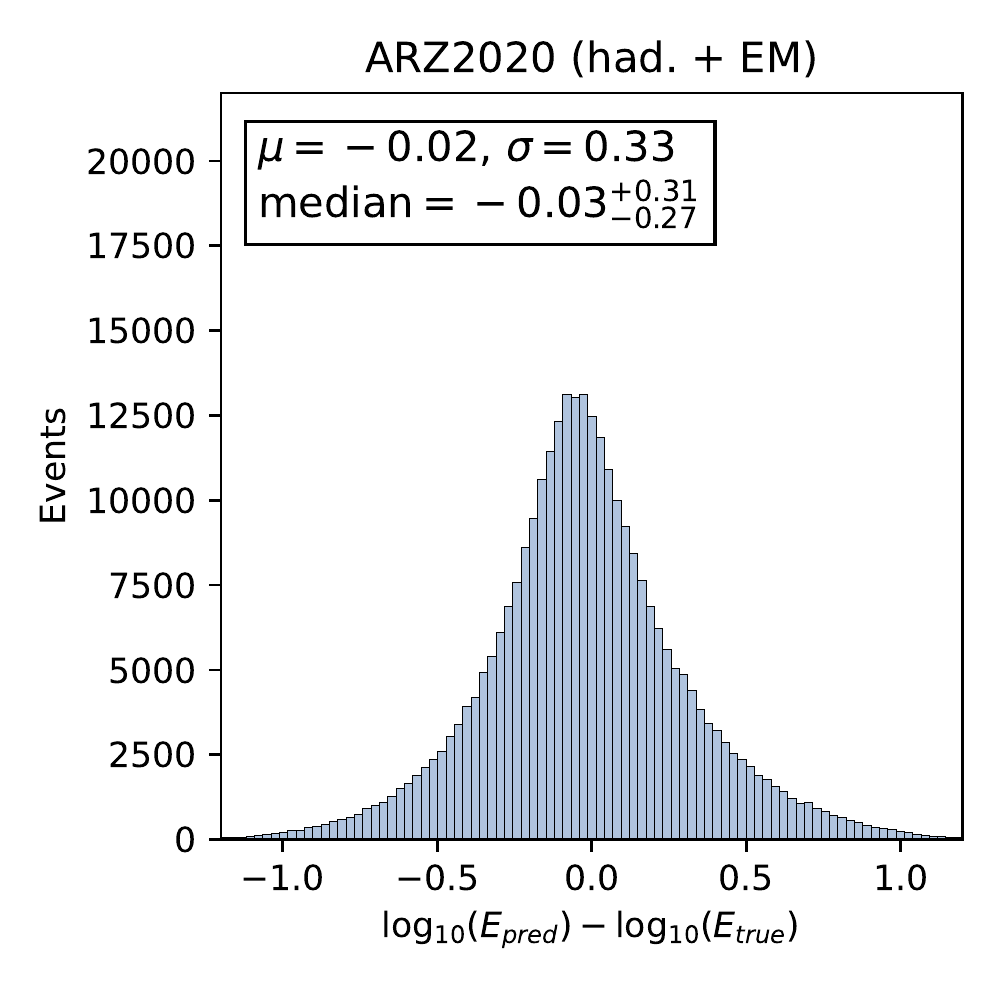} 
    
    \caption{Energy difference histograms for the 3 datasets.}
    \label{fig:energy_normal_res_energy_histogram}
\end{figure*}

After the neural network was trained until convergence, we evaluate the performance on the independent test dataset. We show the resulting energy resolution for all three datasets in Fig.~\ref{fig:energy_normal_res_energy_histogram}. We find similar results for all three datasets with an average standard deviation of approx. 0.3 in the $\log_{10}(E_{\mathrm{shower}})$ which translates to a factor of two on a linear scale. This uncertainty is almost identical to the intrinsic uncertainty from inelasticity fluctuations that we defined as the target resolution. 
The stat boxes in Fig.~\ref{fig:energy_normal_res_energy_histogram} also list the 68\% quantiles around the median which are smaller than the standard deviation, indicating the presence of non-Gaussian tails. These events could potentially be identified and removed through quality cuts in future work. The resolution of the \emph{ARZ2020 (had. + EM)} dataset is only slightly worse (STD = 0.33 vs. 0.31), although the LPM effect and interference with the hadronic shower make this reconstruction more challenging. This result again highlights the usefulness of deep neural networks for such a complex reconstruction task. 

\begin{figure*}[t]
    \centering
    \includegraphics[width=0.28\textwidth, trim={0.10cm 0 0 0}, clip]{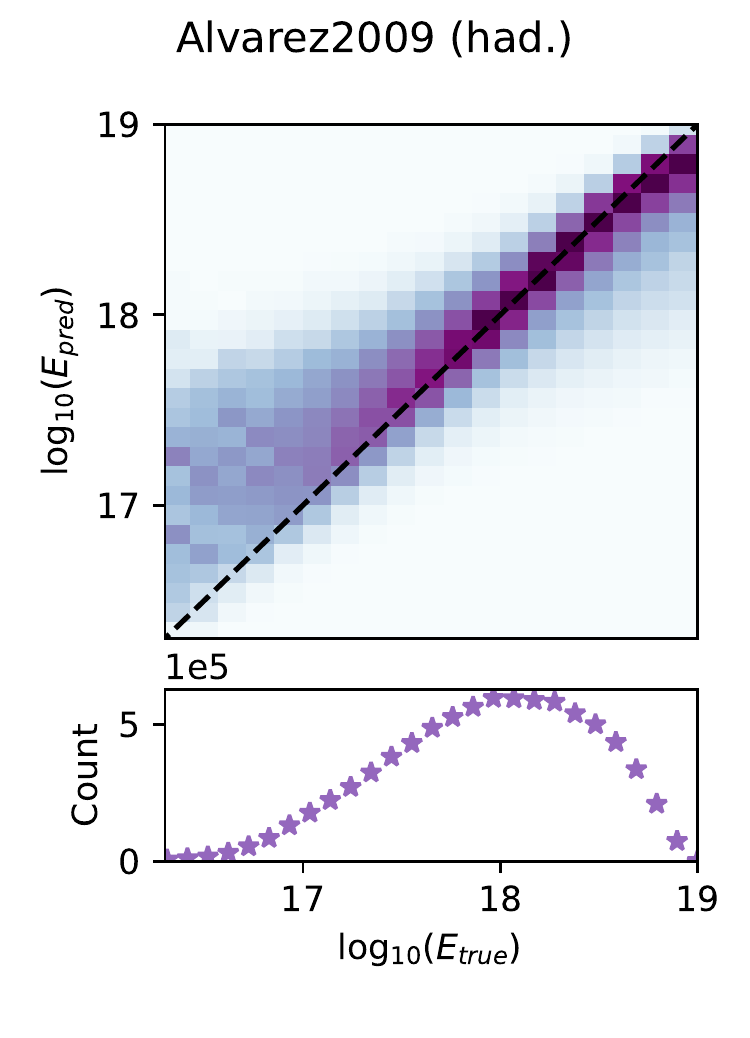}
    \includegraphics[width=0.28\textwidth, trim={0.10cm 0 0 0}, clip]{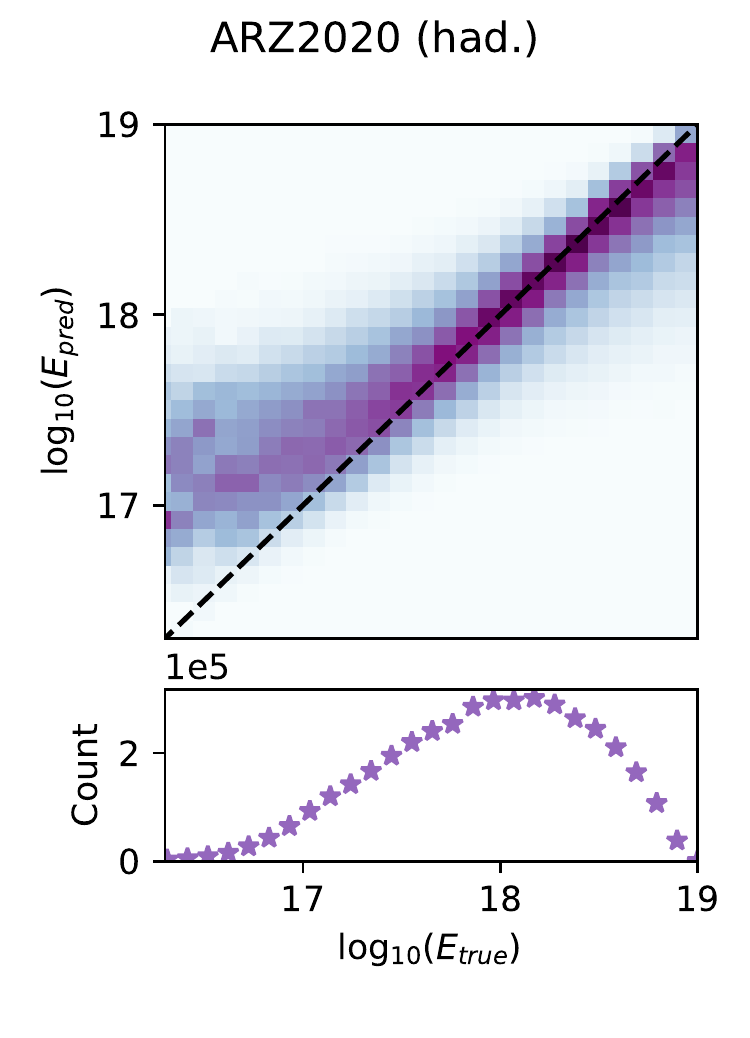}
    \includegraphics[width=0.28\textwidth, trim={0.10cm 0 0 0}, clip]{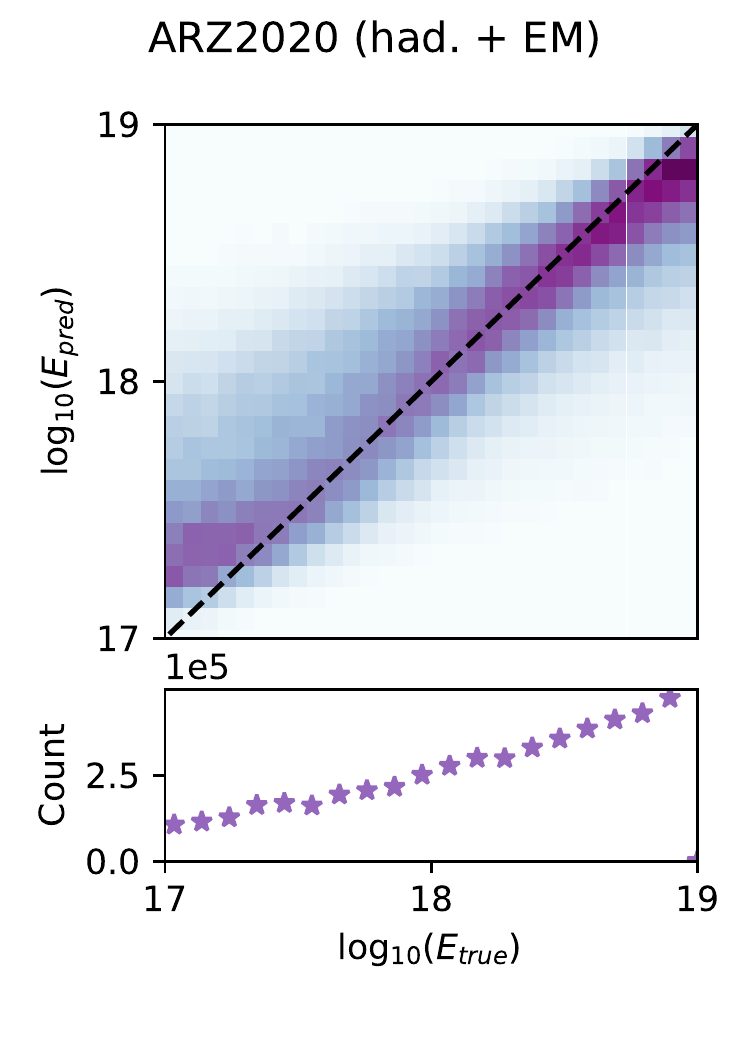} 
    \includegraphics[width=0.08\textwidth, trim={5.5cm 0.5cm 0 1cm}, clip]{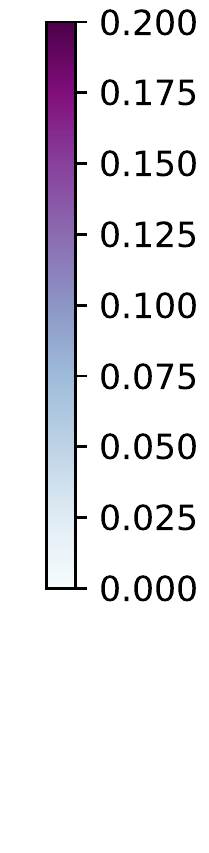} 
    \caption{Predicted energy versus true energy for the 3 datasets. Each column is normalized such that the values sum to unity. The lower panels show the number of events in each energy bin.}
    \label{fig:energy_normal_res_histogram_heatmap}
\end{figure*}

In Fig.~\ref{fig:energy_normal_res_histogram_heatmap} we show the correlation between predicted and true shower energy. All three datasets show a clear correlation. For the \emph{had.} datasets, a significant bias for shower energies below $10^{17.5}~\si{eV}$ is visible. At low energies, the DNN overestimates the true energy. Also at the high-energy end, a systematic underestimation of the energy is visible.
The scatter around the identity line for the \emph{ARZ2020 (had. + EM)} dataset is a bit larger, and a similar bias toward low energies is present but less visible from the figure because of the more restricted shower energy range.

\begin{figure*}[t]
    \centering
    \includegraphics[width=0.9\columnwidth,  clip]{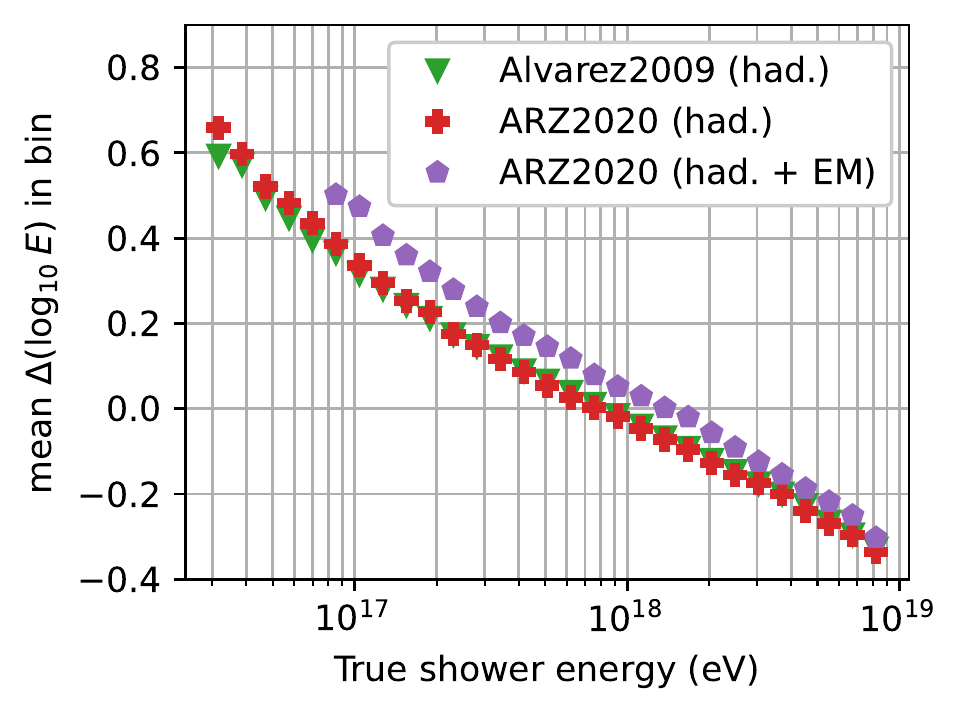}
    \includegraphics[width=0.9\columnwidth, clip]{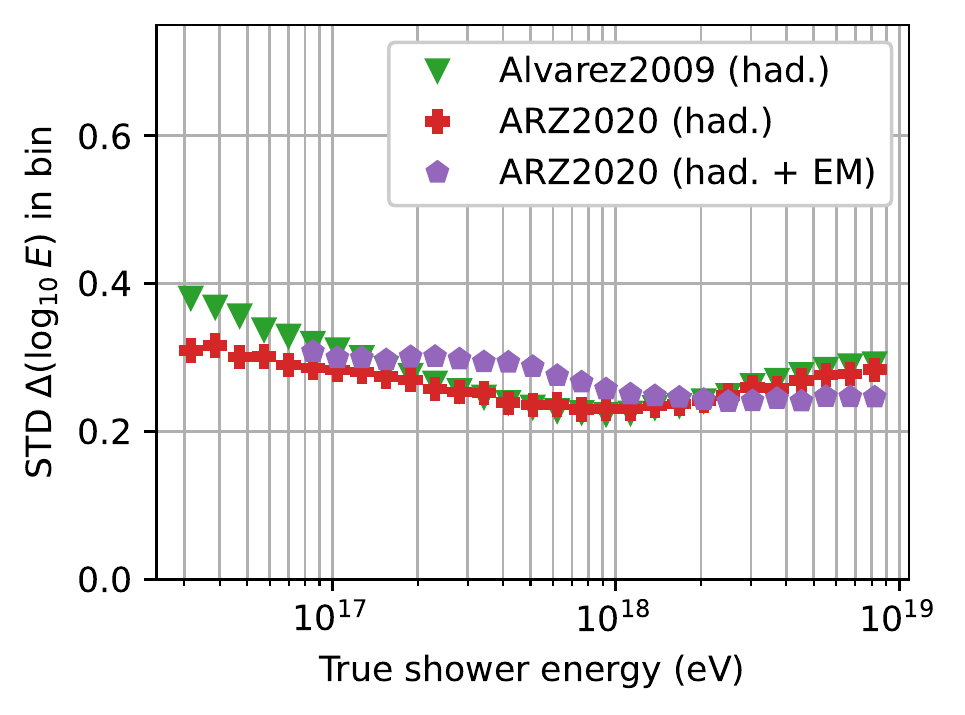} 
    \caption{Mean and STD of the predicted shower energy in bins of true shower energy.}
    \label{fig:energy_normal_res_property_dependence}
\end{figure*}

To gain further insight into the energy dependence, we show the mean and standard deviation in bins of the true shower energy in Fig.~\ref{fig:energy_normal_res_property_dependence}. We find only a weak energy dependence for the standard deviation but a large energy-dependent bias. At low shower energies, the shower energy is overpredicted and at the high-energy end it is underpredicted. In the following, we discuss two approaches to reduce the bias. 

In the data used for training, validating, and testing the neural networks, the events are not distributed uniformly across the energy range, as seen in the lower panels of Fig.~\ref{fig:energy_normal_res_histogram_heatmap}. This will affect the network such that it becomes good at reconstructing events at energies where there are many events in the training dataset, and it will perform worse for events where there are few events. This affects low-energy events the most, as these are the rarest. 
One possible solution to this bias against low-energy events is to use weights in the loss function, such that it puts higher weights on the events that are rarer when compared to more common events. This loss function weight is used for both training and validation. As an example, if a shower energy bin has half the number of events when compared to another shower energy bin in Fig.~\ref{fig:n_events}, then the network will weigh the loss function twice as much for events in that bin as compared to the other bin. This makes it so that the network's weights also take in consideration the less common events, even though they represent a smaller proportion of the training events. We restrict the maximum weighting factors to a range between 1 and 10 to avoid single events dominating a batch. 

\begin{figure*}[t]
    \centering
    \includegraphics[width=0.9\columnwidth,  clip]{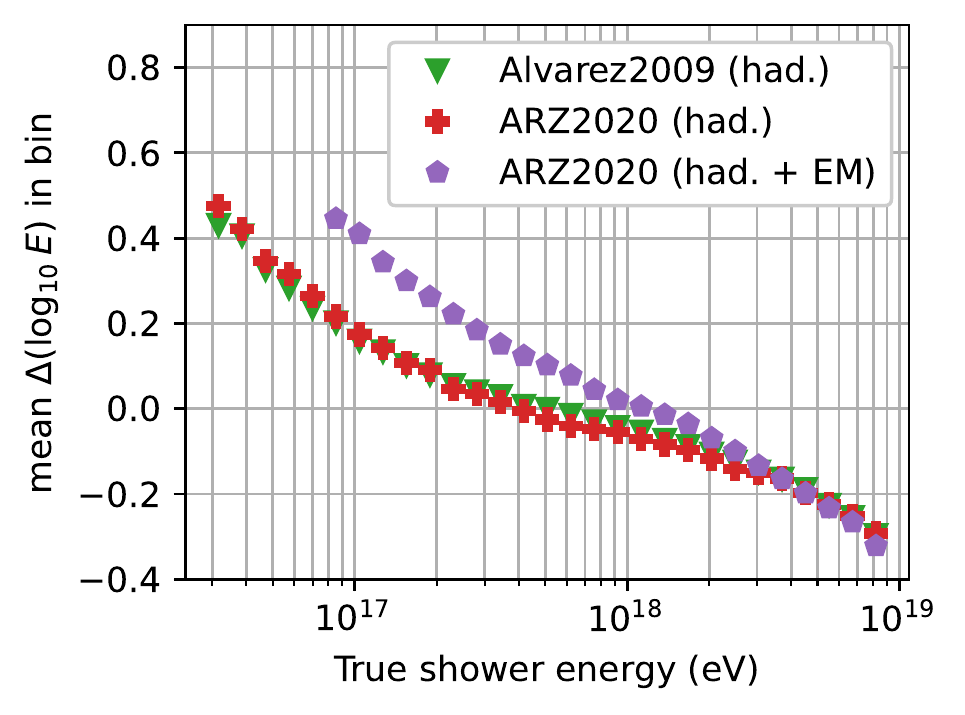}
    \includegraphics[width=0.9\columnwidth, clip]{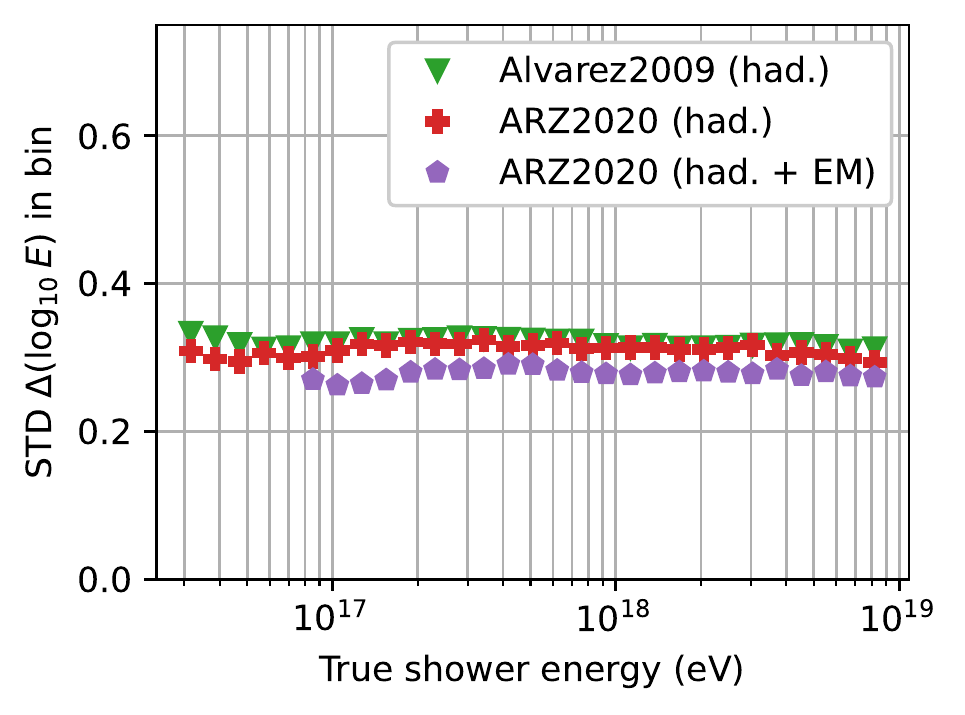} 
    \caption{Mean and STD of the predicted shower energy in bins of true shower energy after re-weighting of the loss function during training with a maximum weight of 10. }
    \label{fig:energy_reweighted_res_property_dependence}
\end{figure*}

Fig.~\ref{fig:energy_reweighted_res_property_dependence} shows equivalent plots to Fig.~\ref{fig:energy_normal_res_property_dependence} for the model that used re-weighting of the loss function during training. It is evident that using re-weighting of the loss function reduces the energy-dependent bias although it does not remove it completely. However, it also makes the network perform slightly worse. This is because the network is now adapting to events that are not as common instead of focusing on the most common events, which leads to the more common events not being reconstructed as well as previously. 
In future work, we recommend making a larger effort for an equal representation of all energies in the datasets to mitigate this effect further. 

\begin{figure*}[t]
    \centering
    \includegraphics[width=0.9\columnwidth,  clip]{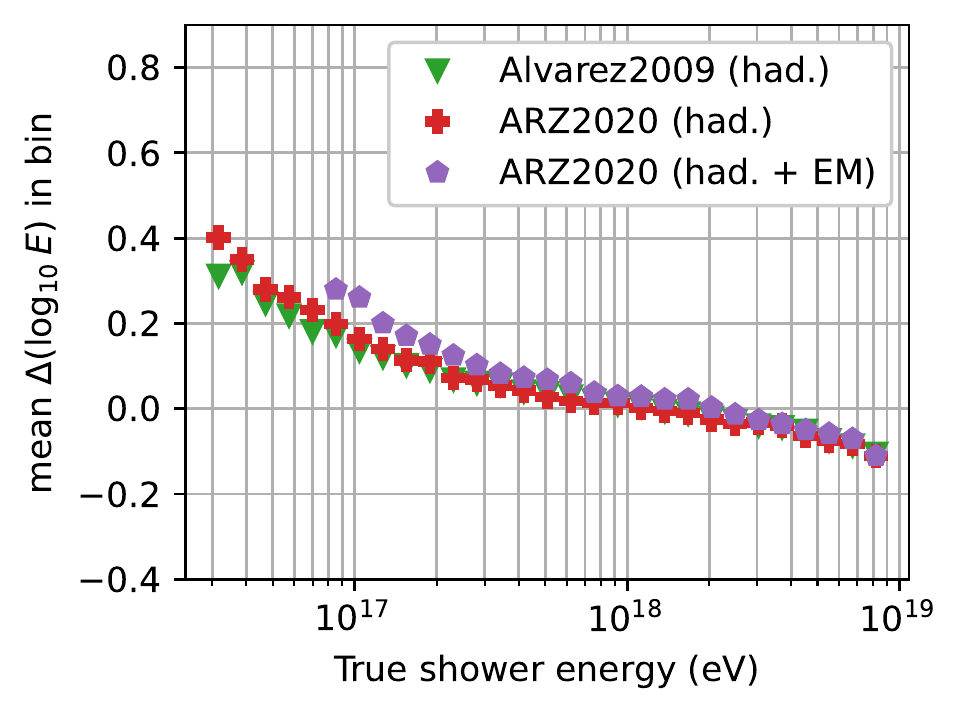}
    \includegraphics[width=0.9\columnwidth, clip]{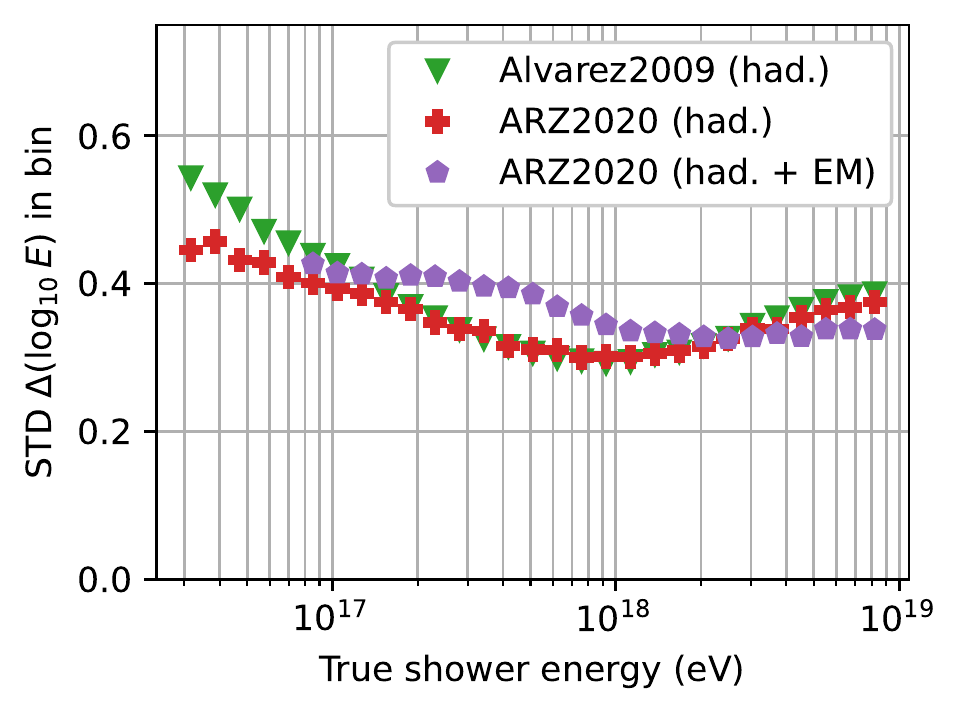}
    \caption{Mean and STD of the predicted shower energy in bins of true shower energy after bias correction (see text for details).}
    \label{fig:energy_biasadjusted_res_property_dependence}
\end{figure*}

Another method to make the bias less significant is to adjust the predicted shower energies for the bias on an event-by-event basis. Once a prediction has been made using the ordinary model, the average bias values at the reconstructed energy from Fig.~\ref{fig:energy_normal_res_property_dependence} are used as an estimate of the bias. This bias is then subtracted from the predicted energy, which yields a bias-adjusted value of the predicted energies. However, because the bias depends on the true shower energy which is unknown, and because the predicted shower energy has significant uncertainties, this bias correction will have significant uncertainties which we see in the resulting distributions shown in Fig.~\ref{fig:energy_biasadjusted_res_property_dependence}. As we expect, the bias is significantly reduced and smaller than what we obtained using the re-weighting approach. However, the uncertainty (standard deviation) is also increased substantially which disfavors using this method.

As in the previous section, we also study how much the reconstruction performance deteriorates if we don't know the event type. The results are presented in Fig.~\ref{fig:energy_event_type} in the appendix. We find that the \ARZem network performs equally well on the \ARZhad dataset than on the \ARZem dataset it was trained on, whereas the \ARZhad network performs worse on the \ARZem dataset. Above \SI{3e18}{eV}, the \ARZem network performs even better on the \ARZhad dataset than the \ARZhad network itself which we attribute to the larger amount of training data at high energies in the \ARZem dataset. At energies below \SI{e18}{eV}, the \ARZhad network still performs better (cf. Fig.~\ref{fig:energy_normal_res_property_dependence}). Hence, there is still a benefit in knowing the event type.

We also repeat our estimation for an upper bound of the systematic uncertainty of the Askaryan emission model by exchanging the \ARZhad and \Alvarez networks/datasets. The results are presented in Fig.~\ref{fig:energy_emission_model} in the appendix. As for the direction reconstruction, we find that the \Alvarez network works better on the \ARZhad dataset than the other way around and attribute it to the same reason.

\section{Summary and Outlook} 
We presented the first end-to-end reconstruction of the neutrino direction and energy from shallow radio detector data using a deep neural network (DNN). The only information available to the DNN was the raw voltage output of the five antenna receivers. The best-performing DNN architecture was a combination of several convolution blocks followed by fully connected layers. The DNN was trained and tested with extensive datasets of several million events of expected radio signals from neutrino interactions with energies between \SI{e17}{eV} and \SI{e19}{eV} at the South Pole. 
Throughout this work we considered all triggered events: i.e., we did not apply any quality cuts with respect to trigger level.

We studied three datasets that differed by how the Askaryan radiation was calculated and by the type of neutrino interaction. To be comparable with previous work, one dataset was restricted to only hadronic showers and used a simple frequency domain parameterization of the Askaryan emission. The second dataset improved on that by using a detailed time-domain calculation of the Askaryan signal. The third dataset considered the challenging event class of electron neutrino charged current $\nu_e$-CC interactions where the shower development is impacted by the LPM effect leading to the interference of multiple particle cascades.
Previous studies often ignored this case or found that the reconstruction performance deteriorated significantly. Our work solved this problem.

The resolution of the neutrino direction (the angular difference between the true and reconstructed direction) shows a narrow peak at $\mathcal{O}$(\SI{1}{\degree}) with extended tails. The point spread function can be described well with a Moffat/King function that can be thought of as a Gaussian with extended tails. We find that the narrow part of the distribution is described with a sigma of \SI{0.6}{\degree} which increases to \SI{0.8}{\degree} for the more challenging $\nu_e$-CC dataset. Due to the tails of the distribution, the 68\% quantiles are at \SI{4}{\degree} and \SI{5}{\degree} respectively. This finding highlights the advantage of a DNN for event reconstruction which is able to learn the complex correlations of $\nu_e$-CC interactions between neutrino direction and observed radio flashes. The performance is only slightly worse than for the simpler non-$\nu_e$-CC interactions which is a significant improvement over previous reconstruction methods. We find the angular resolution improves with both neutrino energy and signal-to-noise ratio. The energy dependence is partly due to an underrepresentation of low-energy events in the training dataset. 

We also find promising results for the energy reconstruction. 
The DNN achieved an average resolution of a factor of 2 in shower energy ($\sigma \approx$ 0.3 in $\log_{10}(E)$) for all triggered events: i.e., without applying any quality cuts with respect to trigger level. The reconstruction uncertainty matches the science requirements of UHE neutrino detectors and is below the intrinsic uncertainty from inelasticity fluctuations. The reconstruction also works surprisingly well for the more complicated $\nu_e$-CC interactions. We find a bias toward the low-energy end, where the DNN overpredicts the shower energy. This is again partly due to the underrepresentation of these events in the training dataset which we plan to address in future work.

Although these are promising results, this work is just the beginning of exploiting deep learning for event reconstruction of radio detector data. Additional network architectures can be exploited to potentially improve the reconstruction performance such as recurrent neural networks for encoding the time domain waveforms. Furthermore, we plan to extend the DNN with a prediction of the event-by-event uncertainty which is needed for many physics analyses. The simplest way to achieve that would be to assume a Gaussian distribution of the uncertainty and to let the network predict the corresponding sigma parameter. While Gaussian errors are an acceptable approximation of the energy uncertainty, the expected angular uncertainties are not Gaussian but banana-shaped regions in the sky \cite{GlaserICRC2019}. In future work, we will use \emph{normalizing flows} \cite{papamakarios2021normalizing} to directly predict the event-by-event probability distribution instead of a single direction plus uncertainty \cite{Gluesenkamp2020}.

Another critical aspect is the accuracy of the training data. With the development of NuRadioMC \cite{NuRadioMC} and its ongoing improvements (see e.g., \cite{Garcia2020, ICRC2021Leptons, ICRC2021RadioPropa, Heyer2022}), a lot of work has already been done, but more work is needed to ensure the dataset's validity: i.e., that it correctly resembles real data. Our study of the impact of different Askaryan emission models already explored a part of potential systematic uncertainties of the MC dataset, but more work is needed to systematically study uncertainties of the simulation code as well as to test robustness against experimental uncertainties. 

The work presented here also opens up new ways to optimize future detectors. Currently, optimization with respect to the reconstruction performance is often not possible as it takes too much time to adapt traditional reconstruction techniques to changing detector layouts. With deep learning based reconstruction, the network architecture can be tuned once, and then retrained on different detector designs quickly to evaluate the reconstruction performance as long as enough training data can be provided. Additional efforts to exchange the time-consuming MC simulation with fast surrogate models -- e.g., by using generative adversarial networks in combination with the reconstruction methods developed here -- would allow the construction of a fully-differentiable pipeline that enables an end-to-end optimization of the detector design: e.g., the position and orientations of the antennas of a shallow radio detector station \cite{ModeWhitepaper2022}. 

\section{Acknowledgments}
We thank all developers of the NuRadioMC code for enabling the creation of the training datasets. We thank the anonymous reviewers, Jakob Beise and Beth Riley for the feedback on the manuscript.
The computations and data handling were enabled by resources provided by the Swedish National Infrastructure for Computing (SNIC) at UPPMAX partially funded by the Swedish Research Council through grant agreement no. 2018-05973.
The work of SM and PB in part supported by NSF grant NRT 1633631 to PB.

% \section*{Appendix}
\begin{figure*}[t]
    \centering
    \includegraphics[width=0.9\columnwidth, clip]{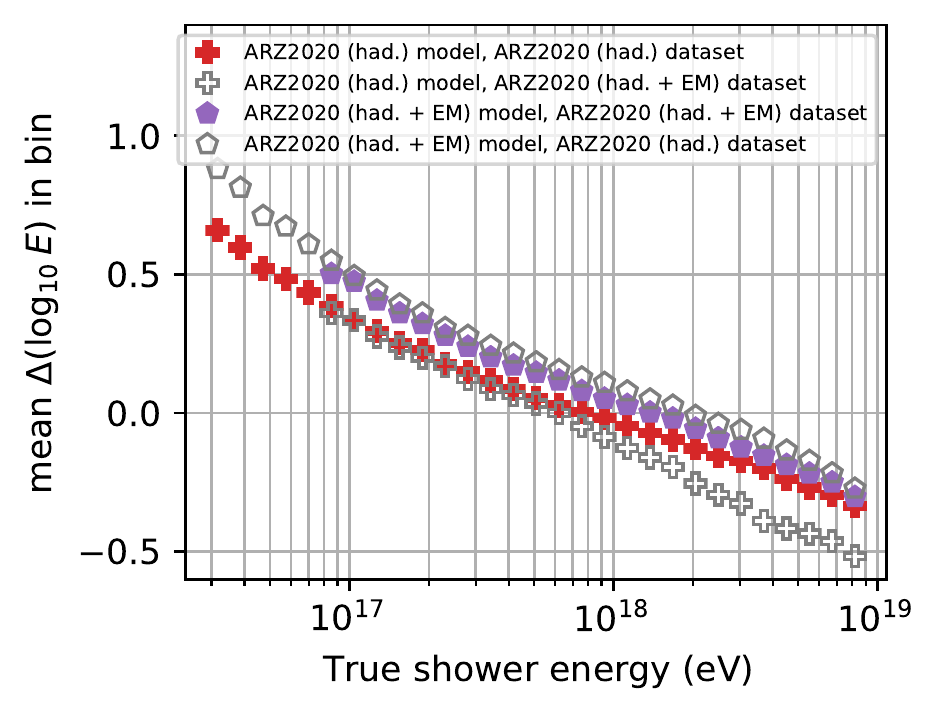}
    \includegraphics[width=0.9\columnwidth, clip]{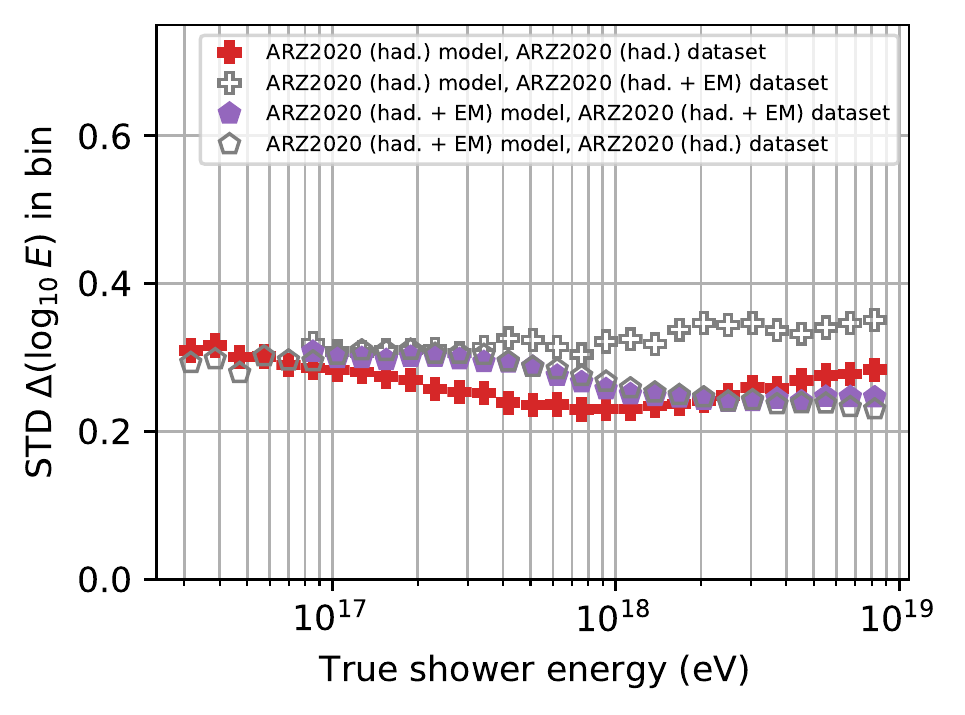}
    \caption{Energy resolution (bias left, standard deviation right) as a function of neutrino energy for an incorrect guess of the event type, i.e., the neural network trained on $\nu_e$-CC events is evaluated on the non-$\nu_e$-CC dataset and vice versa. Filled markers show the correct combinations, empty markers the incorrect combinations.}
    \label{fig:energy_event_type}
\end{figure*}

\begin{figure*}[t]
    \centering
    \includegraphics[width=0.9\columnwidth, clip]{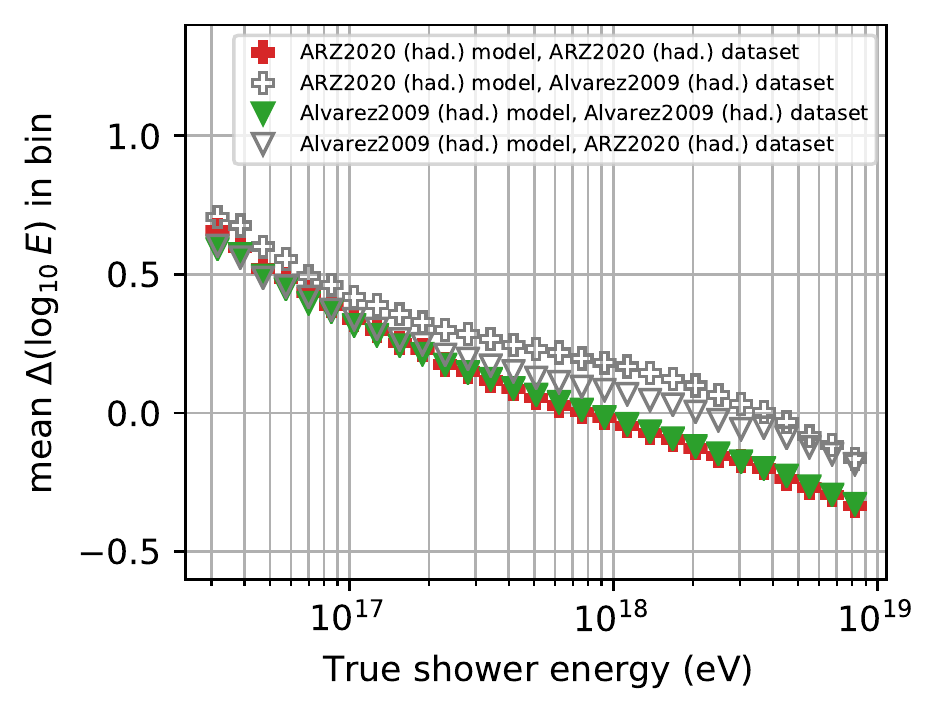}
    \includegraphics[width=0.9\columnwidth, clip]{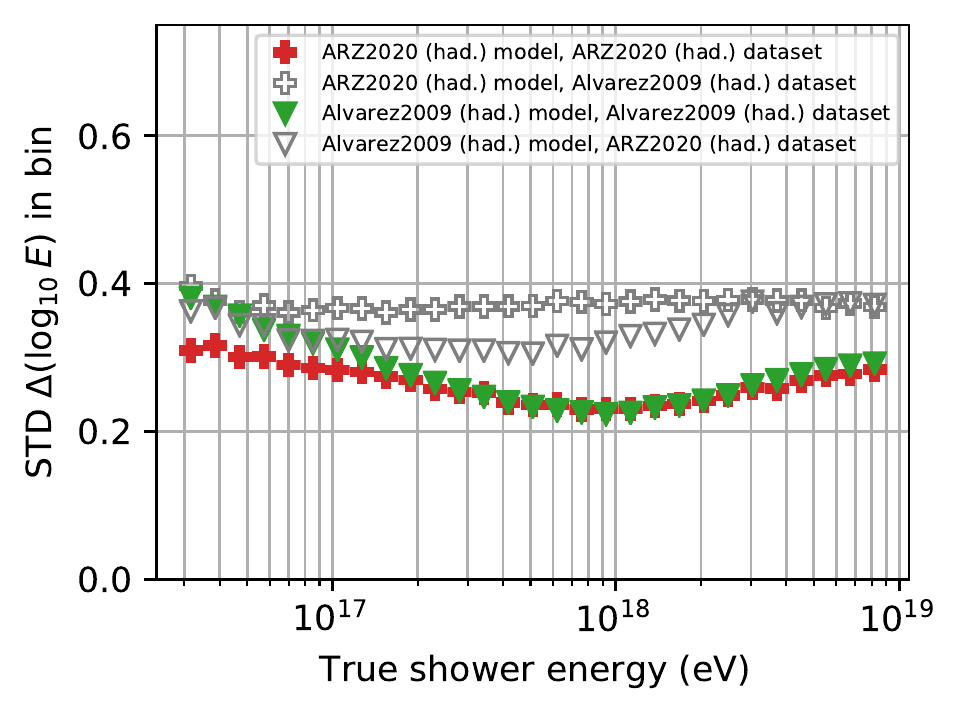} 
    \caption{Energy resolution (bias left, standard deviation right) as a function of neutrino energy for an incorrect combination of the Askaryan emission model, i.e., the neural network trained on the \Alvarez dataset is evaluated with the \ARZhad dataset and vice versa. Filled markers show the correct combinations, empty markers the incorrect combinations.}
    \label{fig:energy_emission_model}
\end{figure*}

\bibliographystyle{JHEP}
\bibliography{bib}
\end{document}